\documentclass[sigconf]{acmart}

\usepackage{graphicx}
\usepackage{colortbl}
\usepackage[T1]{fontenc}
\usepackage{amsmath}
\usepackage{natbib}
\usepackage{amsfonts}
\usepackage[noend]{algorithmic}
\usepackage[nameinlink,capitalize]{cleveref}
\usepackage{float}
\usepackage{algorithm}
\usepackage{subcaption}
\usepackage{dsfont}
\usepackage{multirow}
\usepackage{array}
\usepackage{enumitem}
\usepackage{nccmath}
\setlist[itemize]{leftmargin=4mm}
\usepackage[colorinlistoftodos,textsize=tiny]{todonotes}
\usepackage{multibib}
\newcites{AP}{Appendix References}

\usepackage{soul}
\usepackage{xspace}
\definecolor{navy}{rgb}{0.1, 0.1, 0.8}
\definecolor[named]{gray}{rgb}{0.4, 0.4, 0.4}
\definecolor[named]{olive}{rgb}{0.1, 0.5, 0.1}
\definecolor[named]{ruby}{rgb}{0.8, 0.1, 0.3}
\definecolor{darkpastelgreen}{rgb}{0.01, 0.75, 0.24}
\definecolor{celestialblue}{rgb}{0.29, 0.59, 0.82}
\definecolor{coral}{rgb}{1.0, 0.5, 0.31}
\definecolor{Goldenrod}{rgb}{0.8,0.8,0}

\newcommand{\eat}[1]{}
\newcommand{\rev}[1]{{#1}}

\newcommand{\revA}[1]{{#1}}
\newcommand{\verify}[1]{#1}
\newcommand{\verifyK}[1]{{#1}}
\newcommand{\NOTE}[2]{}
\newcommand{\TODO}[2]{}
\newcommand{\nb}[1]{}
\newcommand{\mar}[1]{}

\DeclareMathOperator{\E}{\mathbb{E}}
\DeclareMathOperator{\Prob}{\mathbb{P}}
\DeclareMathOperator{\Real}{\mathbb{R}}

\DeclareMathOperator{\His}{\mathcal{H}}
\DeclareMathOperator*{\argmax}{arg\,max}

\newcommand{\bracket}[1]{\left[#1\right]}

\newcommand{\numberthis}{\addtocounter{equation}{1}\tag{\theequation}}

\def\BibTeX{{\rm B\kern-.05em{\sc i\kern-.025em b}\kern-.08emT\kern-.1667em\lower.7ex\hbox{E}\kern-.125emX}}

\newcolumntype{L}[1]{>{\raggedright\let\newline\\\arraybackslash\hspace{0pt}}m{#1}}
\newcolumntype{C}[1]{>{\centering\let\newline\\\arraybackslash\hspace{0pt}}m{#1}}
\newcolumntype{R}[1]{>{\raggedleft\let\newline\\\arraybackslash\hspace{0pt}}m{#1}}
\copyrightyear{2020} 
\acmYear{2020} 
\setcopyright{acmcopyright}\acmConference[CIKM '20]{Proceedings of the 29th ACM International Conference on Information and Knowledge Management}{October 19--23, 2020}{Virtual Event, Ireland}
\acmBooktitle{Proceedings of the 29th ACM International Conference on Information and Knowledge Management (CIKM '20), October 19--23, 2020, Virtual Event, Ireland}
\acmPrice{15.00}
\acmDOI{10.1145/3340531.3411861}
\acmISBN{978-1-4503-6859-9/20/10}

\begin{document}
\fancyhead{}

\newcommand{\titlename}{\verify{Describing and Predicting Online Items with Reshare Cascades via Dual Mixture Self-exciting Processes}}

\title{\titlename}

\author{Quyu Kong}
\affiliation{%
  \institution{Australian National University \&\\ UTS \& Data61, CSIRO}
  \city{Canberra}
  \country{Australia}}
\email{quyu.kong@anu.edu.au}

\author{Marian-Andrei Rizoiu}
\affiliation{%
  \institution{University of Technology Sydney \& Data61, CSIRO}
  \city{Sydney}
  \country{Australia}
}
\email{marian-andrei.rizoiu@uts.edu.au}

\author{Lexing Xie}
\affiliation{%
 \institution{Australian National University \& Data61, CSIRO}
 \city{Canberra}
  \country{Australia}}
  \email{lexing.xie@anu.edu.au}

\begin{abstract}
    It is well-known that online behavior is long-tailed, with most cascaded actions being short and a few being very long. A prominent drawback in generative models for online events is the inability to describe unpopular items well.
    This work addresses \revA{these} shortcomings by proposing dual mixture self-exciting processes to jointly learn from groups of cascades.
    \revA{We first start from the observation that maximum likelihood estimates for content virality and \rev{influence} decay are separable in a Hawkes process.} %
    \revA{Next,} our proposed model, which leverages a Borel mixture model and a kernel mixture model, \revA{jointly models the unfolding of a heterogeneous set of cascades.}
	\revA{When applied to} cascades \rev{of} the same online items, the model \revA{directly characterizes their spread dynamics and} supplies interpretable quantities, such \rev{as} content virality and content influence decay, as well as methods for predicting the final content popularities.
    On two retweet cascade datasets \revA{--- one relating to YouTube videos and the second relating to controversial news \rev{articles} ---} we show that our models capture the differences between online items at the granularity of items, publishers and categories.
    In particular, we are able to distinguish between far-right, conspiracy, controversial and reputable online news articles based on how \rev{they} diffuse through social media, achieving an F1 score of 0.945.
    On holdout datasets, we show that \revA{the dual mixture model} provides, for reshare diffusion cascades especially unpopular ones, better generalization performance and, for online items, \revA{accurate \rev{item} popularity predictions}.
\end{abstract}

\maketitle

\newcommand*{\acth}{{\verifyK{\textit{ActiveRT2017-Fit}}}\xspace}
\newcommand*{\actc}{{\verifyK{\textit{ActiveRT2017-Test}}}\xspace}
\newcommand*{\fakeh}{{\verifyK{\textit{RNCNIX-Fit}}}\xspace}
\newcommand*{\fakec}{{\verifyK{\textit{RNCNIX-Test}}}\xspace}

\section{Introduction}

\revA{Online social media platforms disseminate a wide array of content, such as news} articles, photos and videos. 
\revA{For instance, it is common for users to tweet about YouTube videos they enjoy, which are in turn retweeted by their followers},
resulting in \emph{diffusion cascades} of reshares. 
The \revA{amount of reshares that} an item attracts \revA{on the social media platform can consistently influence the total attention that the item receives, also} defined \revA{as its} \emph{popularity}.
\revA{Not all content is made equal and, intuitively, the capacity to command reshare cascades in social media and their characteristics are informative of the content's type, publisher or even veracity (say for online news).
In this work, we characterize online items based on how they are shared and diffused through online social media.}

\revA{When studying what makes diffusion cascades popular,} a \rev{family} of point process based models, \rev{known as} the Hawkes processes, has attracted growing attention~\citep{Zhao2015SEISMIC:Popularity,Mishra2016FeaturePrediction}. Most modeling efforts \revA{concentrate on learning from popular diffusions, usually discarding} unpopular ones. For instance, \citet{Zhao2015SEISMIC:Popularity} only study cascades with at least $50$ retweets.
However, \revA{to characterize online items it is not feasible to apply the same cascade-level filtering, as} all online items \revA{generate}
both ``successful'' and ``unsuccessful'' diffusions. In fact, the latter makes up for a large portion of all cascades, even for popular items, due to the long-tailed distributions~\citep{goel2012structure}.

In this work, we address two open questions \revA{relating to characterizing online items using their social media reshare cascades.}

\begin{figure}[!tbp]
	\centering
	\includegraphics[width=0.48\textwidth]{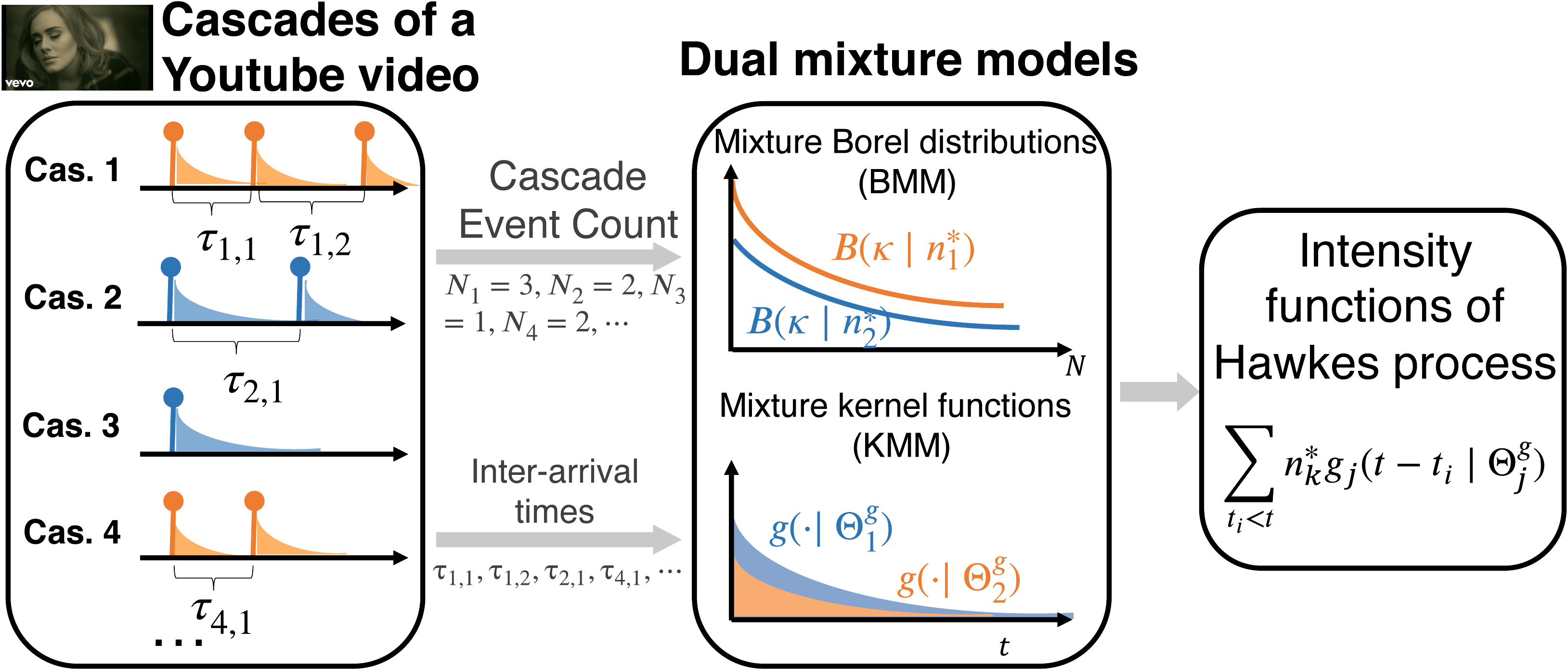}
	\caption{
		Given a group of cascades relating to an online item (e.g., a YouTube video), the dual mixture model fits \revA{separately} a Borel mixture model (BMM) on the cascade event counts, and a kernel mixture model (KMM) on the inter-arrival times. 
		\revA{Finally}, the fitted BMM and KMM are combined to construct the Hawkes intensity functions.
	}
	\label{fig:true_teaser}
	\vspace{-4mm}
\end{figure}

\revA{The first open question relates to jointly modeling a group of heterogeneous cascades \rev{of} the same item.}
Popularities are known to be hard to predict, whether one uses discriminative predictors~\citep{cheng2014can} or generative models~\citep{Rizoiu2017c}.
This suggests that learning from \emph{popular} diffusions \revA{on an online item} leads to modeling bias \revA{as it omits the dynamics of unpopular cascades.} %
The question is: %
\textbf{what representations can account for the diffusions of an online item, as a collection of popular and unpopular cascades?}
We answer this question in two steps. 
\revA{First, we adopt a new representation for Hawkes point processes that decouples content virality and influence decay (i.e., the decaying of influence from a reshare action).
As a result, we find that the maximum likelihood estimates of model parameters are also separable, leading to a \rev{seprate} learning over multiple cascades.}
\revA{In the second step, we propose a novel dual mixture self-exciting model that captures the diverse diffusion dynamics that each online item encounters across a set of cascades.}
One is a Borel mixture model~\citep{daw2018queue} \revA{that accounts} for the distribution of final \rev{reshare counts} \revA{for each \rev{cascade} in the set}, and the other is a kernel mixture model \revA{that controls the inter-arrival time dynamics}
for capturing influence decaying dynamics. \rev{\cref{fig:true_teaser} illustrates the model where a group of cascades (left) relating to a video is modeled by the dual mixture model (middle) and the two fitted models are combined to form the intensity function of the mixture processes (right).}

\revA{The second open question is}
\textbf{how can we apply the mixture models to describe online items and predict final content popularities?}
Fitted \revA{model} parameters and derived quantities are commonly used for analyzing individual reshare cascades~\citep{Mishra2016FeaturePrediction}.
One can describe an online item or a content producer by compiling the key parameters of the dual mixture models. In this work, we build quantities that summarize \revA{respectively} an item's content virality and influence decay. We also construct \emph{diffusion embeddings} that describe the item/producer reshare dynamics and
can be used with off-the-shelf supervised and unsupervised tools.
\revA{We deploy our methods}
on two large-scale retweet cascade datasets, \revA{the first} about YouTube videos and \revA{the second around far-right, conspiracy, controversial and reputable} online news articles.
\revA{When using the diffusion embeddings, we find that content producers group together with respect to video category and publisher virality \rev{in the YouTube dataset}.
For the news dataset,} the publishers of reputable and controversial news form two separable \revA{clusters, and we obtain} an F1 score of $0.945$ \revA{when using a \rev{Gradient Boosting Machine} to distinguish the two types of news.}
To accurately predict the final popularities \revA{of newly posted items},
\revA{we fit our proposed dual mixture model on the historical information --- how previously posted items spread --- and we leverage it for}
recent items. 
On \revA{both} datasets, we \revA{show that for individual cascades the dual mixture models provide} improved generalization performance compared to \revA{individual cascade fits}~\citep{Mishra2016FeaturePrediction} and non-mixture models, especially for unpopular cascades. \revA{For online items,} \rev{the model} obtains the best final popularity prediction \revA{when compared against} feature-based regressors and the state-of-the-art generative models~\citep{Zhao2015SEISMIC:Popularity,kong2019modeling}.

The main contributions of this work are:
\begin{itemize}
    \item \revA{\textbf{\rev{Separable} joint learning.} 
    We adopt a new representation for Hawkes processes that separates virality and influence decay, which leads to a \rev{separable learning} of model parameters \rev{in} the maximum likelihood estimates}
	\item \revA{\textbf{Dual mixture self-exciting processes.}
	We design mixture models for the two separable model factors --- a Borel mixture for the virality, and a kernel mixture for the influence decay --- in order to capture the diverse diffusion dynamics that each online item encounters across a set of cascades.}
    \item \revA{\textbf{Item characterization and item popularity prediction.}
    We propose a set of tools to quantify online items using their spread dynamics: derived quantities, and the diffusion embeddings. We also propose methods for predicting item final popularities.}
    \item \revA{\textbf{Two real-world case studies.}}
    On two large retweet datasets, we \revA{show our \rev{methods are} effective for the unsupervised exploratory analysis of collections of online publishers, and in predicting content category --- for example whether a news article is controversial}.
	We \revA{also} show better generalization and popularity prediction performances for unseen items.
\end{itemize}

\section{Related Work}
Generative models are commonly employed for modeling temporal diffusions of online information. 
Such models are designed to predict final popularities \citep{Zhao2015SEISMIC:Popularity,bao2016modeling}, uncover hidden diffusion networks \citep{rodriguez2011uncovering} and detect rumors \citep{ma2016detecting}. 
\revA{Feature-driven models predict popularity by}
training machine learning algorithms \revA{using} 
statistical summaries of \revA{resharing events} together with user features and content features~\citep{bakshy2011everyone,martin2016exploring}. 
However, to our knowledge, most of the prior work concentrate on popular cascades, and the complete temporal information of the unpopular diffusions is rarely considered.

Hawkes processes~\citep{hawkes1974cluster} are a class of self-exciting point processes --- past events \revA{spawn} future events --- widely applied in analyzing social media~\citep{Kobayashi2016TiDeH:Dynamics,cao2017deephawkes,zhang2018efficient}, earthquake aftershocks~\citep{ogata1988statistical}, neuronal activity~\citep{apostolopoulou2019mutually}, online advertising~\citep{parmar2017forecasting} and finance~\citep{bacry2015hawkes}. 
The distribution of event counts of Hawkes processes has not been explored until \rev{recently}. While \citet{Rizoiu2017c} and \citet{daw2018queue} are able to obtain the distribution under certain assumptions, \citet{o2020quantifying} show a method to numerically approximate actual event count distributions.
\revA{Our} work \revA{enhances} the understanding of Hawkes processes by connecting its log-likelihood function \rev{with the} event count distribution.

Existing work \revA{leveraging} mixture \revA{with} temporal point processes focuses on two levels. 
Event-level mixture modeling clusters individual events from a sequence~\citep{yang2013mixture,du2015dirichlet}, whereas, as in \revA{our} work, sequence-level mixture modeling identifies clusters of event sequences~\citep{Wu2020discovering}. 
The \revA{prior work most relevant to ours} is \revA{by} \citet{xu2017dirichlet}.
Their model integrates Hawkes processes and a Dirichlet distribution for learning event sequence groups. 
\revA{Our} work extends the prior literature in several ways.
\revA{First,} we derive two separate mixture models from Hawkes processes for modeling content virality and content influence decay separately. 
\revA{Second,} we apply the models to complete historical diffusion cascades for learning and quantifying temporal dynamics of online items.
\section{Preliminaries}\label{sec:preliminary}
In this section, we first define diffusion cascades. 
Next, we introduce the Hawkes processes, together with essential concepts including its cluster representation, branching factor, size distribution and likelihood function.

\noindent\textbf{Diffusion cascades.}
In online social media platforms, such as Twitter, users read content posted by others, and they can \rev{reshare} it, exposing the content to \rev{a} broader audience. 
The initial posting event and the following \rev{reshare} events together constitute a diffusion \textit{cascade}.
In this work, we analyze groups of cascades that discuss about \rev{the} same online items, e.g., an online video~\citep{Rizoiu2016ExpectingPopularity}, an image meme~\citep{lakkaraju2013s}, or a news article~\citep{tan2014effect}.
Mathematically, we denote a cascade $i$ discussing an online item $v$ as $\His_{v, i} = \{t_0, t_1, t_2, \dots, t_{N_{v,i}-1}\}$ where $N_{v,i} \geq 1$ is the number of events in cascade $i$ of item $v$, $\forall t_j \in \His_{v, i}$ are event times on $[0, \infty)$ relative to $t_0$ and $t_0 = 0$ is the initial event time. 
Let $\His_{v, i}(T), N_{v, i}(T)$ represent the event set and the event count before time $T$, respectively, i.e., $\His_{v, i}(T) = \{ t_j \mid t_j \in \His_{v,i}, t_j < T\}$ and $N_{v,i}(T) = |\His_{v,i}(T)|$. The total event count $N_{v, i}$ is also known as the \textit{popularity} of the cascade $i$. The \textit{popularity} of the online item $v$ is then the total popularity of all related cascades.

\noindent\textbf{Hawkes processes} are special classes of self-exciting point processes in which the occurrence of new events will increase the likelihood of future event happening~\citep{hawkes1971spectra}. 
In Hawkes processes, the event intensity is a function conditioned on the past occurred events and we choose the intensity function in a form similar to~\citep{Zhao2015SEISMIC:Popularity}:
\begin{align} \label{eq:intens}
    \lambda(t \mid \His_{i}(t)) = \mu + \sum_{t_j \in \His_{i}(t)} n^* g(t - t_j)
\end{align}
where $\mu$ is the background event rate, $n^*$ is known as the \textit{branching factor}, $g: \Real^+ \rightarrow \Real^+$ is a memory kernel encoding the time-decaying influence of past events on future events and $\int_0^{\infty}g(\tau) d\tau = 1$.
While~\cref{eq:intens} is equivalent to existing definitions of Hawkes processes~\citep{hawkes1971spectra,laub2015hawkes}, it explicitly incorporates $n^*$ as a model parameter which simplifies derivations in~\cref{sec:joint_learning}.
We note that for information cascades (such as retweet cascades on Twitter), there is no background intensity, as all the retweets are considered to be \rev{spawned by} the original tweet, i.e., $\mu = 0$.
Common choices of the memory kernels include the exponential kernel function~\citep{xu2016learning}, $g_{EXP}(\tau) = \theta e^{-\theta \tau}$, 
the power-law kernel~\citep{Mishra2016FeaturePrediction}, $g_{PL}(\tau) = \theta c^{\theta} (\tau + c)^{-(1+\theta)}$, among others. 
\rev{We} refer to \citep{kong2019modeling} for a review of kernels used with cascades.

\begin{figure}[!tbp]
	\centering
	\includegraphics[width=0.45\textwidth]{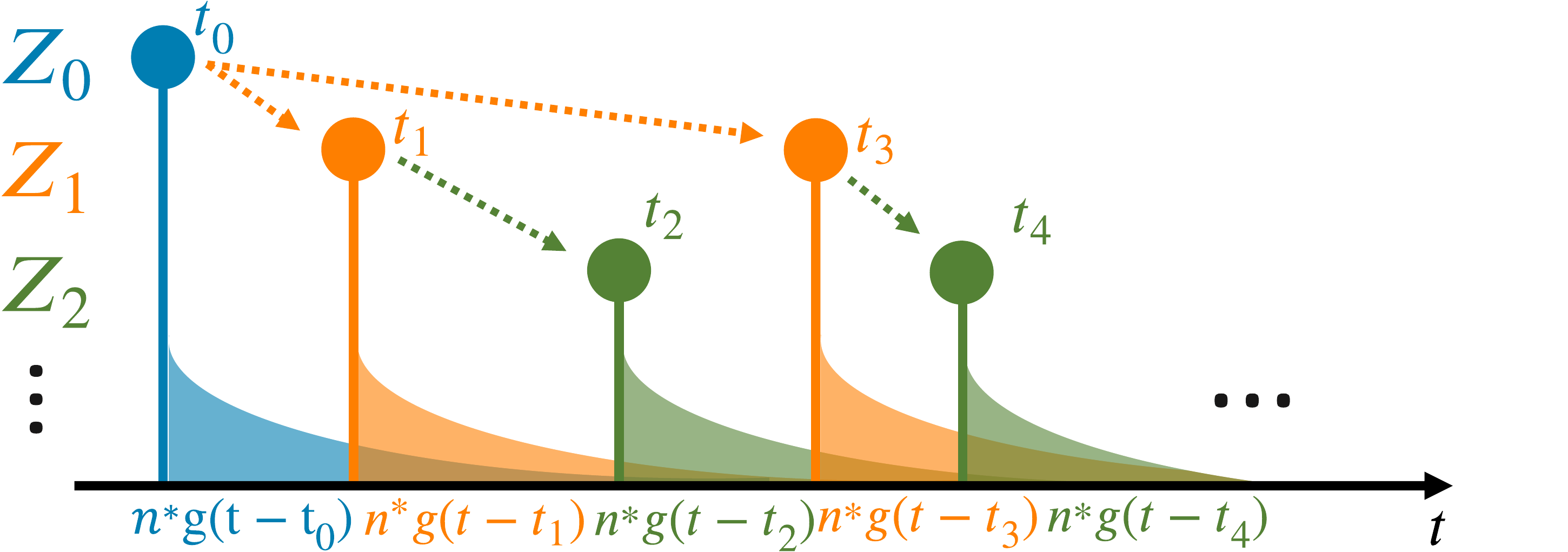}
	\caption{
		The cluster representation of a Hawkes process. 
		Each individual event $t_i$ initiates an inhomogeneous Poisson process with the intensity function $n^* g(t-t_i)$ (identical for all events). 
		Different generations of events are shown in distinct colors; arrows indicate the parent-offspring relation; and the event counts at each generation form a branching process, i.e., $\{Z_0, Z_1, Z_3, \dots\}$.
	}
	\label{fig:teaser}
	\vspace{-4mm}
\end{figure}

\noindent\textbf{Cluster representation and size distribution.}
An alternate representation of the Hawkes self-exciting process is a latent cluster of Poisson processes, introduced by \citet{hawkes1974cluster}.
\cref{fig:teaser} depicts the cluster representation of an example Hawkes process, with highlighted parent-offspring relations between events. 
Each event generates offspring events following an inhomogeneous Poisson process with the intensity function $n^* g(t)$, which means its number of offspring follows a Poisson distribution of intensity $\int_0^T n^* g(t) dt$. 
When $T \rightarrow \infty$, the event counts at each generation --- denoted as $\{Z_0, Z_1, Z_2, \dots\}$ --- produce a Galton-Watson branching process whose offspring distribution is a Poisson distribution with intensity $n^*$~\citep{durrett2019probability}.
The total size of a Hawkes process can be then computed as $N = \sum_{n} Z_n$. 
This quantity is known as the \emph{total progeny number} of the branching process, following a Borel distribution~\citep{borel1942emploi}, denoted as
$%
    \mathbb{B}(\kappa \mid n^*) = \Prob [N = \kappa \mid n^*] = \frac{(\kappa n^*)^{\kappa-1} e^{-\kappa n^*}}{\kappa!}
$%
, which holds for $n^* < 1$.
The mean and variance of a Borel distribution are $\frac{1}{1-n^*}$ and $\frac{n^*}{(1-n^*)^3}$.
The analysis of Hawkes process size distribution~\citep{o2020quantifying} and this particular analytical form~\citep{daw2018queue} are both very recent developments on the point process literature.

\noindent\textbf{Parameter estimation.} 
The parameters of a Hawkes process can be estimated by maximizing the likelihood function of a general point process~\citep{Daley2008}:
\begin{equation}\label{eq:general_likelihood}
    L(\Theta \mid \His_{i}(T)) = e^{-\int_0^T \lambda(\tau \mid \His_{i}(T)) d\tau}\prod_{t_j \in \His_{i}(T)} \lambda(t_j \mid \His_{i}(T))
\end{equation}

\section{Separable Hawkes processes Fitting} 
\label{sec:joint_learning}
In this section, we discuss jointly learning a single set of parameters from a collection of Hawkes realizations.

Let $\mathbb{H} = \{\His_1, \His_2, \dots\}$ be a set of independent Hawkes realizations, assumed to be generated from the same model parameterized by $n^*$, the branching factor, and $\Theta^g$, the parameter set of $g(\cdot)$.
It is then straightforward to estimate $n^*$ and $\Theta^g$ by maximizing the joint log-likelihood function $\mathcal{L}(n^*, \Theta^g \mid \His)$ defined as the sum of the individual log-likelihoods (i.e., the log of \cref{eq:general_likelihood}):
\begin{equation}\label{eq:sum_likelihood}
    \mathcal{L}(n^*, \Theta^g \mid \mathbb{H}) = \sum_{\His_i \in \mathbb{H}} \log L(n^*, \Theta^g \mid \His_i)
\end{equation}

After plugging~\cref{eq:general_likelihood} into~\cref{eq:sum_likelihood}, we see that the joint log-likelihood function can be rearranged as a sum of two functions with independent parameter sets given $\int_0^{\infty}g(\tau) d\tau = 1$ and $T \rightarrow \infty$ (detailed in the online appendix~\citep{appendix}):
\begin{align}\label{eq:sum-two-likelihoods}
    \mathcal{L}(n^*, \Theta^g \mid \mathbb{H}) = \mathcal{L}_g(\Theta^g \mid \mathbb{H}) + \mathcal{L}_n( n^* \mid \mathbb{H})
\end{align}
\revA{with} $\mathcal{L}_g$ a function of $\Theta^g$ and $\mathcal{L}_n$ a function of $n^*$:

\begin{align}
    \mathcal{L}_g(\Theta^g \mid \mathbb{H}) &= \sum_{\His_i \in \mathbb{H}} \sum_{t_j \in \His_i, j \geq 1}\log \sum_{t_z < t_j} g(t_j - t_z \mid \Theta^g) \label{eq:likelihood_g} \\
    \mathcal{L}_n( n^* \mid \mathbb{H}) &= \sum_{\His_i \in \mathbb{H}} \log \bracket{(n^*)^{N_i - 1} e^{-N_i n^*}} \label{eq:likelihood-n*}
\end{align}
Regarding the assumption $T \rightarrow \infty$, we show in~\cref{sec:experiment} that most cascades are complete in practice given a large $T$. We also note that \cref{eq:likelihood-n*} can be solved efficiently and analytically by setting its first derivative to $0$.

The above results indicate that $\Theta^g$ and $n^*$ can be learned independently in
two separate phases, by maximizing $\mathcal{L}_g$ and $\mathcal{L}_n$.
This amounts to fitting $n^*$ from observed final cascade sizes only, and $\Theta^g$ from inter-arrival times between events.

We note that maximizing $\mathcal{L}_n$ is equivalent to the maximum likelihood estimation of the Borel distribution. One can see this by expanding both forms, as shown below:
\begin{align*}
    &\hspace{-5mm}\argmax_{n^*} \sum_{\His_i \in \mathbb{H}} \log \mathbb{B} (N_i \mid n^*) \\
    &= \argmax_{n^*} \sum_{\His_i \in \mathbb{H}} \bracket{\log (n^*)^{N_i - 1} e^{-N_i n^*} + \log \frac{N_i^{N_i-1}}{N_i !}} \\
    &\stackrel{\text{(a)}}{=} \argmax_{n^*} \mathcal{L}_n( n^* \mid \mathbb{H}) \numberthis
\end{align*}
where we discard the \revA{log ratio of} constants $N_i$ at step (a).

To the best of our knowledge, this is the first work to discuss the \revA{separable} form of Hawkes parameter estimations and its connection to the Borel distribution.

\section{Dual Mixture Model}\label{sec:uncovering}
\revA{In practice, an online item is reshared across a set of diffusion cascades of diverse dynamics.}
In this section, we \revA{propose a dual mixture model that allows individual cascades to differ one from another.}
\revA{Given the} separability \revA{of the} log-likelihood functions (\cref{eq:likelihood_g,eq:likelihood-n*}),
we introduce a Borel mixture model (BMM) and a kernel mixture model (KMM) to automatically uncover the latent clusters of models based on cascade sizes and time intervals.
\revA{Finally, we employ the fitted dual mixture model to construct item level characterizations, such as $\hat{n}^*_v$, $\hat{\theta}_v$ and the diffusion embeddings \rev{with a distance measure}.}

\noindent \textbf{\revA{Mixture models for Hawkes processes.}}
We are given $\mathbb{H}_v$, a set of cascades relating to an online item $v$, and the number of components $k_v$ --- \revA{there exist} $k_v$ latent generative models with unknown relations to the cascades in $\mathbb{H}_v$.
We seek to learn $k_v$ groups of $n^*$ and $\Theta^g$, and their weights.
As indicated in~\cref{sec:joint_learning}, we model these two parameter sets separately using cascade sizes and inter-arrival times.
We denote the obtained model as $M_v = \{M^B_v, M^K_v\}$ where $M^B_v = \{(n^*_1, p^B_1), \dots, (n^*_{k_v}, p^B_{k_v})\}$, $M^g_v = \{(\Theta^g_1, p^g_1), \dots, (\Theta^g_{k_v}, p^g_{k_v})$.
$p^B_1, \dots, p^B_{k_v}$ and $p^g_1, \dots, p^g_{k_v}$ are the component weights for corresponding Borel models and kernel functions.

Given two mixture models, $M^B_v$ and $M^K_v$, inferred separately from a group of cascades, we assume the intensity functions of the corresponding Hawkes processes --- \cref{eq:intens} --- are parameterized by the cartesian product of $M^B_v$ and $M^K_v$, i.e.,
\begin{equation}\label{eq:mixture_to_hawkes}
    M^H_v = \{(n^*_i, \Theta^g_j, p^B_i p^g_j) \mid (n^*_i, p^B_i) \in M^B_v \text{ and } (\Theta^g_j, p^g_j) \in M^g_v \}
\end{equation}
where $p^B_i p^g_j$ gives the component weight. \cref{fig:true_teaser} summarizes the modeling procedure.

\noindent {\bf Borel mixture model (BMM).} 
To learn the $M^B_v$ for the online item $v$, we present an EM estimation algorithm~\citep{dempster1977maximum}.
A BMM can be fitted on $\mathbb{H}_v$ by maximizing the log-likelihood
\begin{equation} \label{eq:BMM_likelihood}
    \mathcal{L}_{BMM} = \sum_{\His_{v,i} \in \mathbb{H}_v} \log \sum_{k=1}^{k_v} \underbrace{p^B_k \mathbb{B}(N_{v,i} \mid n^*_k)}_{q^B(k, N_{v,i})}
\end{equation}
As maximizing~\cref{eq:BMM_likelihood} directly suffers from the identifiability issue~\citep{bishop2006pattern}, we apply the Expectation-Maximization (EM) algorithm commonly used for learning mixture models~\citep{tomasi2004estimating}. This algorithm optimizes an alternative lower bound $Q_{BMM}$ defined as
\begin{equation}
    Q_{BMM} = \sum_{\His_{v,i} \in \mathbb{H}_v}  \sum_{k=1}^{k_v} p^B(k\mid N_{v,i}) \log q^B(k, N_{v,i})
\end{equation}
where $p^B(k \mid N_{v,i})$ is the probability of $N_i$ being a member of the $k$th model and is updated during the E step. Next we give the update formulas for the E and M steps.

E-step: membership probabilities are updated
\begin{equation}
    p^B(k \mid N_{v,i}) = \frac{q^B(k, N_{v,i})}{\sum_{j=1}^{k_v} q^B(j, N_{v,i})}
\end{equation}

M-step: $n^*_k$ and $p^B_k$ are updated analytically
\begin{align}
    &(n^*_k)^{new} = \frac{\sum_{N_{v,i}}p^B(k\mid N_{v,i})(N_{v,i} - 1)}{\sum_{N_{v,i}}p^B(k\mid N_{v,i})N_{v,i}} \\ 
    &(p^B_k)^{new} = \sum_{N_{v,i}}\frac{p^B(k\mid N_{v,i})}{|\mathbb{H}_v|}
\end{align}
Parameters are updated iteratively by alternating these two steps until the convergence of $\mathcal{L}_{BMM}$.

\noindent {\bf Kernel mixture model (KMM).}
As we follow similar derivations for obtaining $M^K_v$, we note only two differences regarding the definition of $\mathcal{L}_{KMM}$ and the update of $\Theta^g_k$
\begin{align}
    &\mathcal{L}_{KMM} = \sum_{\His_{v,i} \in \mathbb{H}_v} \log \sum_{k=1}^{k_v} p^g_k f^g( \His_{v,i} \mid \Theta^g_k) \label{eq:l_kmm} \\
    &(\Theta^g_k)^{new} = \argmax_{\Theta^g}  \sum_{\His_{v,i} \in \mathbb{H}_v} p^g(k\mid \His_{v,i}) \log f^g( \His_{v,i} \mid \Theta^g) \nonumber
\end{align}
where $f^g(\His_{v,i} \mid \Theta^g) = \prod_{t_j \in \His_{v,i}}\sum_{t_z < t_j} g(t_j-t_z\mid \Theta^g)$. The way $(\Theta^g_k)^{new}$ is solved depends on specific kernel functions. In our experiments, we solve this with a non-linear solver, Ipopt~\citep{Wachter2006}, where a power-law kernel function is employed.

\cref{eq:BMM_likelihood,eq:l_kmm} \revA{have respectively} linear and quadratic computational complexity, \revA{however} the EM algorithm \revA{allows an} efficient \revA{implementation of} the dual mixture model.
Detailed derivations of the BMM and the KMM can be found in the online appendix~\citep{appendix}.

\noindent {\bf Determining \revA{the number of} components.} 
\revA{Prior literature uses a number of}
information criteria for choosing a component number of mixture models~\citep{burnham2004multimodel, lukovciene2009determining}, including the Akaike information criteria (AIC). 
In our experiments, we employ AIC defined as $2k_v - 2\mathcal{L}_{BMM}$ to select $k_v$ with BMMs. 
\revA{Note that fitting BMM is computationally efficient}
--- due to the analytical updates of the EM algorithm --- \revA{which} allows one to experiment various values for $k_v$. 
In our experiments, the numbers of components $k_v$ given by AIC are generally \revA{between} $2$ \revA{and} $5$.

\noindent {\bf \revA{Characterizing items using the dual mixture model.}}
\revA{We build item-level quantifications based on the dual mixture model fitted on all cascades relating to the given item.
The diffusion embedding provides a fixed length vector describing the information in the components of BMM and KMM, while the content virality and influence decay provide single value summarizations of the two mixtures.}

A diffusion embedding constructed from the fitted mixture models $M_v$ is a vector of mixture component weights. Taking the power-law kernel function as an example, we build a diffusion embedding in two steps:
\begin{itemize}
    \item Parameter discretization: we first discretize the continuous model parameters $n^*$, $\theta$ and $c$ by separating them into fixed number of quantile bins. 
    Given BMMs learned from all observed online items $V$, we obtain the value of the $i$th quantile $q^{n^*}_i$ from the weighted samples $\{(n^*_j, p^B_{j}) \mid j \in \{1,\dots, k_v\}, \forall v \in V \}$. We use the algorithm provided in~\citep{harrell2017hmisc} to compute weighted quantiles. Similarity, we get $q^{c}_i$, $q^{\theta}_i$ from the fitted KMMs.
    \item Weight aggregation: we then convert $M^B_v$ into a vector of weights for an online item $v$, $\pmb{m}^{n^*}_v = [m^{n^*}_{v,1}, \dots]^\mathsf{T}$ where each element is the sum of weights $m^B_{v,i} = \sum_{q^{n^*}_{i-1} < n^*_j \leq q^{n^*}_i} p^B_j$. Moreover,  $M^g_v$ can be encoded as $\pmb{m}^{c}_v = [m^{c}_{v,1}, \dots]^\mathsf{T}$ and $\pmb{m}^{\theta}_v = [m^{\theta}_{v,1}, \dots]^\mathsf{T}$.
\end{itemize}
In the end, three vectors ($\pmb{m}^{n^*}_v, \pmb{m}^{c}_v, \pmb{m}^{\theta}_v$) are provided for each online item as the diffusion embeddings and can be used with off-the-shelf supervised or unsupervised tools.

We also compute the single value summarizations as:
$%
    \hat{n}^*_v = \sum_{k=1}^{k_v} n^*_k p^B_k, \hat{c}_v = \sum_{k=1}^{k_v} c_k p^g_k, \hat{\theta}_v = \sum_{k=1}^{k_v} \theta_k p^g_k
$. %
We denote $\hat{n}^*_v$ as content virality, and $\hat{\theta}_v$ as influence decay.
\revA{These are two values of interest showing} how viral and how long the influence of an online item stay in online discussions.

\noindent {\bf Distance \revA{between} diffusion embeddings.} 
\revA{Given two items described by their respective} 
diffusion embeddings $(\pmb{m}^{n^*}_1, \pmb{m}^{c}_1, \pmb{m}^{\theta}_1)$ and $(\pmb{m}^{n^*}_2, \pmb{m}^{c}_2, \pmb{m}^{\theta}_2)$, we seek to measure their distance $D_{1,2}$.
We note that the position of elements in the embeddings represents quantiles at an increasing order, but common distance measures, such as the Euclidean distance and the cosine distance, ignore such information. For example, given $\pmb{m}^{n^*}_{1} = [1,0,0, \cdots]$, $\pmb{m}^{n^*}_{2} = [0, 1, 0, \cdots]$ and $\pmb{m}^{n^*}_{3} = [0, 0, 1, \cdots]$, $\pmb{m}^{n^*}_{1}$ is intuitively closer to $\pmb{m}^{n^*}_{2}$ than to $\pmb{m}^{n^*}_{3}$ instead of equally close. To address this, we employ the Wasserstein distance~\citep{arjovsky2017wasserstein} which accounts for positional information. The Wasserstein distance of order 1 for single dimensional histogram has a closed-form solution defined as 
$%
    W_1 (\pmb{M}^{n^*}_{1}, \pmb{M}^{n^*}_{2}) = \sum_{i} |M^{n^*}_{1, i} - M^{n^*}_{2, i}|
$, %
where $\pmb{M}^{n^*}_{\cdot} = [\sum_{j=1}^1 \pmb{m}^{n^*}_{\cdot,j}, \sum_{j=1}^{2} \pmb{m}^{n^*}_{\cdot,j}, \sum_{j=1}^{3} \pmb{m}^{n^*}_{\cdot,j}, \cdots]$ represents the cumulative weights at increasing quantiles. We then define the distance of the pair of diffusion embeddings as
\begin{align} \label{eq:embedding_distance}
    D_{1, 2} = W_1 (\pmb{M}^{n^*}_{1}, \pmb{M}^{n^*}_{2}) + W_1 (\pmb{M}^{c}_{1}, \pmb{M}^{c}_{2}) + W_1 (\pmb{M}^{\theta}_{1}, \pmb{M}^{\theta}_{2})
\end{align}

\section{Predicting the future of cascades}
\label{sec:prediction}

In this section, we show how fitted mixture models can be applied to future observations. We describe the evaluation of generalization performance on holdout parts of unseen cascades. Next, we derive predictions of final popularities. %

\noindent {\bf Models for future content.} We build mixture models for a newly published item by combining historical fitted models of items $V_{\rho}$ from the same publisher $\rho$, i.e.,
\begin{align}
    M^B_{\rho} &= \bigcup_{v \in V_{\rho}} \{(n^*_i, p^B_i/|V_{\rho}|), \cdots\}, \hspace{5mm}\forall (n^*_i, p^B_i) \in M^B_v\\
    M^g_{\rho} &= \bigcup_{v \in V_{\rho}} \{(\Theta^g_i, p^g_i/|V_{\rho}|), \cdots\}, \hspace{5mm}\forall (\Theta^g_i, p^g_i) \in M^B_v
\end{align}
and $M_{\rho} = \{M^B_{\rho}, M^g_{\rho}\}$, assuming the new item follows the dynamics of its predecessors. 
Following~\cref{eq:mixture_to_hawkes}, we obtain $M^H_{\rho}$ from $M_{\rho}$. 
In our experiments, \rev{we limit $V_{\rho}$ to the most recent published items.}

\noindent
{\bf Cascade holdout log-likelihood.} When fitting a Hawkes process on a cascade $\His_i(T)$ until an observation time $T$, the log-likelihood value of the holdout part of this cascade, i.e., $HLL = \mathcal{L}(\Theta \mid \His_i) - \mathcal{L}(\Theta \mid \His_i(T))$, evaluates the model generalization performance to unseen events. For our proposed \revA{dual} mixture model, we compute an expected holdout log-likelihood stemming from the posterior model probabilities given $\His_i(T)$, i.e.,
\begin{align} \label{eq:expected_hll}
    \E \bracket{HLL} = \sum_{(n^*_k, \Theta^g_j, p^B_k p^g_j) \in M^H_{\rho}} & \bracket{\mathcal{L}(\Theta \mid \His_i) - \mathcal{L}(\Theta \mid \His_i(T))} \times \nonumber \\
    &\hspace{2mm}\Prob[n^*_k, \Theta^g_j \mid \His_i(T)]
\end{align}
where \revA{we have}:
$%
    \Prob[n^*_k, \Theta^g_j \mid \His_i(T)] = \frac{\Prob[\His_i(T) \mid n^*_k, \Theta^g_j]p^B_k p^g_j}{\sum_{M^H_{\rho}} \Prob[\His_i(T) \mid n^*, \Theta^g]p^B p^g} 
$%
\begin{table*}[tbp]
	\centering
	\setlength{\tabcolsep}{4pt}
	\caption{Statistics of the two social media datasets.}
	\vspace{-1mm}
	\begin{tabular}{rrrrrrrrr}
	  \toprule
	  & Start time & End time & \#categories & \#publishers & \#items & \#cascades & \#tweets \\ 
	  \midrule
		\acth{} &  Jan 1, 2017 & May 1, 2017 & \multirow{2}{*}{\shortstack[r]{$18$ (\textit{Music},\\ \textit{Gaming}, ...)}}& \multirow{2}{*}{$11,297$ channels} & \multirow{2}{*}{$75,717$ videos} &  \multirow{2}{*}{$30,535,891$} & \multirow{2}{*}{$85,334,424$} \\ 
		\actc{} & Jun 1, 2017 & Dec 31, 2017 & & &  & & \\ 
        \fakeh{} & June 30, 2017 & Jan 1, 2019 & \multirow{2}{*}{\shortstack[r]{$2$ (\textit{RNIX},\\ \textit{CNIX})}} &  \multirow{2}{*}{$73$ domains} & \multirow{2}{*}{$102,429$ articles} & \multirow{2}{*}{$8,129,126$} & \multirow{2}{*}{$56,397,252$}  \\ 
		\fakec{} & Feb 1, 2019 & Dec 31, 2019 &  &  & & & \\
		\bottomrule
	\end{tabular}
	\label{tab:dataset-profiling}
	\vspace{-2mm}
\end{table*}

\noindent
{\bf Cascade posterior size distribution.} 
Given a pair of parameters $n^*$ and $\Theta^g$, we are able to derive the posterior size distribution given $\His_i(T)$ of a cascade $i$. 
The future events after time $T$ are of two kinds: direct offspring of observed events (their count denoted as $N^d_i$) and indirect offspring (children of children, total count denoted as $N^{ind}_i$). 
The process generating direct offspring is an inhomogeneous Poisson process of conditional intensity $\lambda(t | \His_i(T)), t > T$ --- note that this is not a stochastic function as only the history up to time $T$ is accounted in the intensity function.
Consequently, $N^d_i$ follows a Poisson distribution of the intensity $\Lambda_i(T \mid n^*,  \Theta^g) = \int_T^{\infty} \lambda(\tau | \His_i(T), n^*, \Theta^g) d\tau$.
Furthermore, each direct offspring initiated a Hawkes process and its total progeny number follows a Borel distribution. 
Given the number of direct offspring $N^d_i$, the total number of direct and indirect offspring follows
a Borel-Tanner distribution (also known as the generalized Borel distribution)~\citep{Haight1960}:
$
    \mathbb{B}(\kappa \mid n^*, N^d_i) = \frac{N^d_i(\kappa n^*)^{\kappa -N^d_i} e^{-\kappa n^*}}{\kappa (\kappa-N^d_i)!}
$
for $\kappa = N^d_i, N^d_i +1, \cdots$. Its mean, $\frac{N^d_i}{1-n^*}$, and variance, $\frac{N^d_i n^*}{(1-n^*)^3}$, are similar to those of a Borel distribution.

Finally, the posterior cascade size distribution is therefore
\begin{align}\label{eq:posterior_distribution}
    &\Prob[N_i = n \mid \His_i(T)] = N_i(T) \\
    &\hspace{7mm}+ \sum_{z = 0}^{n-N_i(T)} Poi(z \mid \Lambda_i(T \mid n^*,  \Theta^g)) \mathbb{B} (n-N_i(T) \mid n^*, z) \nonumber
\end{align}
where $Poi(\cdot | \lambda)$ is the Poisson distribution given intensity $\lambda$. 
\cref{eq:posterior_distribution} leads to a quadratic complexity in computing the final size distribution, which is intractable in most real-life scenarios.
A numerical trick can be applied to reduce the complexity by introducing a threshold probability $\epsilon_p$ and summing until $Poi(z \mid \Lambda_i(T \mid n^*,  \Theta^g)) < \epsilon_p$.

\noindent
{\bf Online item popularity prediction.} The final popularity of an online item consists of two parts in prediction: the final popularities of current observed cascades and new cascades created in future.

We first use past average cascade counts of the publisher $\rho$ as an estimation of \revA{the} new cascades \revA{that will} emerge in future, denoted as $\hat{C}_{\rho}$. The final popularity of these is thus the mean of a Borel-Tanner distribution given $\hat{C}_{\rho}$ initial events, i.e.,$\frac{\hat{C}_{\rho}}{1-n^*}$. We then compute the mean values from a posterior distribution as the predicted final popularity $\hat{N}_{v,i}$ of the observed cascade $i$ given $n^*$ and $\Theta^g$, i.e.,
\useshortskip
\begin{align*}
    \hat{N}_{v,i}(n^*, \Theta^g) &\\
    &\hspace{-6mm} = N_{v,i}(T) + \sum_{\kappa = 0}^{\infty}\sum_{z = 0}^{\kappa} \kappa \cdot Poi(z \mid \Lambda_{i}(T \mid n^*,  \Theta^g)) \mathbb{B} (\kappa \mid n^*, z) \\
    &\hspace{-6mm}\stackrel{\text{(a)}}{=} N_{v,i}(T) +\sum_{z = 0}^{\infty}  Poi(z \mid \Lambda_{i}(T \mid n^*,  \Theta^g))  \sum_{\kappa = z}^{\infty} \kappa \cdot \mathbb{B} (\kappa \mid n^*, z) \\
    &\hspace{-6mm}\stackrel{\text{(b)}}{=} N_{v,i}(T) + \frac{\sum_{z = 0}^{\infty} z \cdot Poi(z \mid \Lambda_{i}(T \mid n^*,  \Theta^g))}{1 - n^*} \\
    &\hspace{-6mm}\stackrel{\text{(c)}}{=} N_{v,i}(T) + \frac{\Lambda_{i}(T \mid n^*,  \Theta^g)}{1 - n^*} \numberthis \label{eq:observed-cascade-predicted-size}
\end{align*}
where step (a) exchanges the order of two summations. Step (b) and step (c) follow the means of a Borel-Tanner distribution~\citep{Haight1960} and a Poisson distribution.
Last, we add predictions of all cascades and future cascades relating to a new online item
and take expectation over possible parameter sets from the mixture models
\begin{equation} \label{eq:size_prediction}
    \hat{N}_v = \E_{M^H_{\rho}} \bracket{ \frac{\hat{C}_{\rho}}{1-n^*} + \sum_{\His_{v,i}(T) \in \mathbb{H}_v(T)} \hat{N}_{v,i}(n^*, \Theta^g)}
\end{equation}

As the variance of Borel-Tanner distribution is also known~\citep{Haight1960}, \cref{eq:posterior_distribution} enables us to derive the variance of final popularities.

\section{Experiments and results} \label{sec:experiment}
\revA{This section shows how the proposed dual mixture model is used to characterize online items.
\cref{ssec:datasets} introduces} two \revA{Twitter} datasets used in this work
\revA{and \rev{our experimental} setup.
\Cref{ssec:measurement} analyzes} online items \revA{using} fitted dual mixture models. \revA{In \cref{sec:categorical_prediction}} we \revA{investigate} the predictability of item category.
Finally \revA{in \cref{ssec:future-content}}, we evaluate model generalization and popularity prediction performances on unseen data.~\footnote{The code and dataset can be found at: \url{https://bit.ly/3glRerX}}

\subsection{Datasets}
\label{ssec:datasets}
We conduct experiments on two large-scale retweet cascade datasets concerning the spread of two \revA{types of} online \revA{items} on Twitter: 
YouTube videos (\textit{ActiveRT2017}) and news articles (\textit{RNCNIX}). The content category and publisher information \rev{are} given for each item. 
In the \revA{remainder of this section}, we also explain data collection, preprocessing \revA{and fitting} steps.

\noindent\textbf{\revA{Datasets collection.}}
\revA{The} tweets of \textit{ActiveRT2017} and \textit{RNCNIX} were collected from Twitter public APIs. 
For \textit{ActiveRT2017}, during 2017, tweets mentioning YouTube videos were obtained by querying for YouTube video url handlers (\textit{youtube.com} and \textit{youtu.be}) at the real-time streaming endpoint\footnote{https://developer.twitter.com/en/docs/tweets/filter-realtime/overview}.
The video metadata \revA{was crawled} using a tool provided in \citep{wu2018beyond} including video categories, titles, \revA{textual} descriptions and \revA{the} YouTube channel information \revA{(i.e. the publisher of the video)}. 
We only keep \textit{active} videos where \revA{a video is considered as} \textit{active} \revA{if it} received at least $100$ tweets and $100$ shares within $120$ days~\citep{mishra2019linking}.
\revA{The} tweets of \textit{RNCNIX} were collected by Digital Media Research Centre\footnote{https://research.qut.edu.au/dmrc/}.
They retrospectively queried the Twitter search endpoint\footnote{https://developer.twitter.com/en/docs/tweets/search/overview} \revA{for} tweets mentioning articles from a list of \revA{controversial} news publishers and a list of leading Australian news \rev{outlets}~\citep{bruns2016big,bruns2016publics,bruns2020news}. 
We enriched this dataset with \revA{the textual content} of the news articles. 
For each article, we downloaded the webpage and concatenated the content of \rev{the following} HTML tags --- \textit{<title>}, \textit{<meta name=``description''>}, \textit{<meta name=``keyword''>} and \textit{<p>} --- that generally contain article titles, descriptions, keywords and main text bodies.
\revA{Throughout this paper, we collectively denote YouTube videos (for \textit{ActiveRT2017}) and news articles (for \textit{RNCNIX}) as \emph{online items}.
Similarly, YouTube \rev{channels} (\textit{ActiveRT2017}) and news media website \rev{domains} (\textit{RNCNIX}) are denoted as \emph{publishers}.}
We \rev{keep} online items \rev{that have at least one} cascade with at least $50$ events. 
\revA{Note that} this filtering is different from prior work \revA{as, for selected items, we consider all of their cascades regardless of their \rev{event counts}.}

\noindent\textbf{\revA{Online item category.}}
YouTube video categories are provided \revA{in the} YouTube metadata.
\revA{For \textit{RNCNIX}, we construct two categories (RNIX and CNIX) based on the origin of the publisher:} 
the Reputable News Index (RNIX) \revA{which contains} Australian traditional news media such as \textit{abc.net.au}, and the Controversial News Index (CNIX) \revA{that regroups} news sources that are known for producing controversial news articles (such as \textit{infowars.com} \rev{and} \revA{\textit{breitbart.com}}). 

\noindent\textbf{\revA{Train-test split.}}
We \revA{perform a temporal} split \revA{of each} dataset into two \revA{subsets}: the historical cascades and the test cascades. 
We introduce \revA{a} one month gap between \revA{the two subsets to make sure that all cascades from the \rev{historical} set are finished before the start of the testing set} 
(as discussed in~\cref{sec:joint_learning}). 
We show in the online appendix~\citep{appendix} that \rev{more than} $99 \%$ of \revA{all} cascades \revA{in} our datasets \revA{finish within 30 days}.
\revA{We use the cascades in the \rev{historical} set to fit our dual mixture models, and to produce item-level quantifications.}
\revA{We use the test set to} evaluate model generalization and popularity prediction \revA{on unseen content}.
\cref{tab:dataset-profiling} presents a summary of \revA{the number of items, publishers, cascades and tweets in each} dataset, \revA{together with} the start and end time periods for the \revA{fit-test} splits.

\begin{figure}[!tbp]
	\centering
	\begin{subfigure}{.48\textwidth}
		\includegraphics[width=\textwidth,page=1]{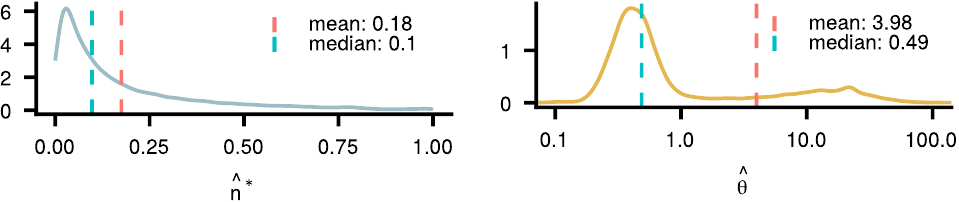}
		\vspace{-6mm}
		\caption{\acth{}}
	\end{subfigure}
	\begin{subfigure}{.48\textwidth}
		\includegraphics[width=\textwidth,page=1]{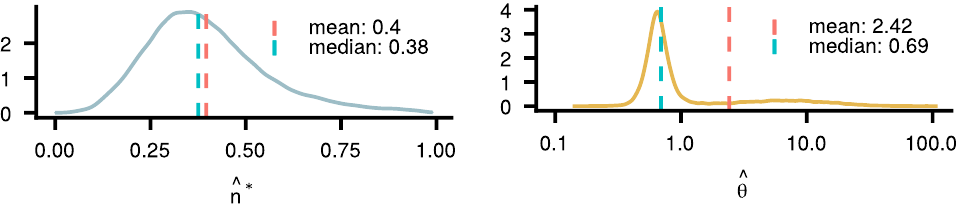}
		\vspace{-6mm}
		\caption{\fakeh{}}
	\end{subfigure}
	\vspace{-4mm}
	\caption{
		Density plots of content \revA{virality} $\hat{n}^*$  of BMMs and content influence decay $\hat{\theta}$ of KMMs fitted on two datasets. Mean and median are shown as red and blue dashed lines.
	}
	\label{fig:param_density}
	\vspace{-2mm}
\end{figure}

\noindent\textbf{\revA{Profiling fitted parameters on training set.}}
\revA{We fit the dual mixture model on all cascades relating to the same online item.}
The number of mixture components \revA{is selected by optimizing the} AIC score for BMM. 
The power-law kernel function is applied for the KMM, as it is shown to outperform others in modeling online information diffusion~\citep{Mishra2016FeaturePrediction}.
\revA{For each item $v$ in each training} dataset \revA{(\acth{} and \fakeh{})}, we obtain the fitted BMM and KMM \revA{parameters, as well as the item-level descriptions introduced in ~\cref{sec:uncovering}:}
$\hat{n}^*_v$, $\hat{c}_v$, $\hat{\theta}_v$, $\pmb{m}^{n^*}_v$, $\pmb{m}^{c}_v$ and $\pmb{m}^{\theta}_v$. 
We set the number of quantiles to $10$ for diffusion embeddings. 
\cref{fig:param_density} \revA{shows the parameters distribution for the item virality $\hat{n}^*$ (left column) and influence decay $\hat{\theta}$ (right column).}
\revA{We see that} \fakeh{} presents higher $n^*$ values and higher $\theta$ values than \acth{} (\revA{visible in the} distributions \revA{as a whole and in the} median values). 
This indicates that news articles \revA{tend to be more viral} than YouTube videos on Twitter, \revA{however they stay for shorter in people's collective memory.
This is expected, given}
the \revA{fast paced} nature of news. Due to space limitation, we also present in the online appendix~\citep{appendix} weighted density plots of fitted BMM and KMM parameters where distributions of different mixture components are shown.

\subsection{Measurements of online items} 
\label{ssec:measurement}
\revA{The parameters of the dual mixture models characterize the online items directly, given that the mixtures are trained on all cascades pertaining to the same item.}
In this section, we \revA{explore the link between item categories and publishers, and the fitted dual mixture models.}

\begin{figure}[!tbp]
    \centering
    \begin{subfigure}{0.35\textwidth}
        \includegraphics[width=\textwidth]{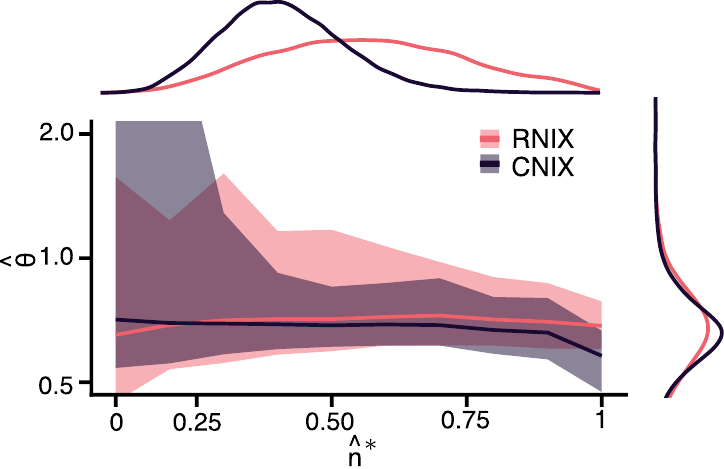}
        \vspace{-3mm}
        \caption{\fakeh{}}
        \label{fig:measurement_b}
    \end{subfigure}
	\begin{subfigure}{0.48\textwidth}
        \includegraphics[width=\textwidth,page=1]{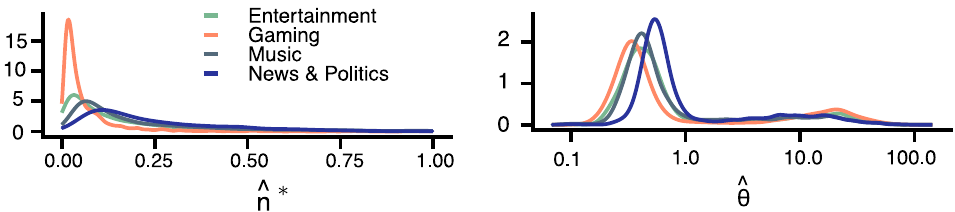}
        \vspace{-3mm}
        \caption{\acth{}}
        \label{fig:measurement_a}
    \end{subfigure}
    \vspace{-3mm}
	\caption{
        Quantify online items at the category level via the aggregated model parameters, $\hat{n}^*$ and $\hat{\theta}$ of two datasets. Fig.(a) \textit{RNIX} and \textit{CNIX} from \fakeh{}: the median and $25^{th}$/$75^{th}$ quantiles of $\hat{\theta}$ (y axis) at varying $\hat{n}^*$ values (x axis) are presented along with densities of $\hat{n}^*$ and $\hat{\theta}$ by sides. Fig.(b) Four popular YouTube video categories, \textit{Music}, \textit{Entertainment}, \textit{Gaming} and \textit{News \& Politics} from \acth{}: density plots of $\hat{n}^*$ and $\hat{\theta}$.
	}
    \label{fig:measurement}
    \vspace{-2mm}
\end{figure}

\noindent {\bf Category-level overview.} 
\revA{First, we investigate whether item categories can be distinguished using virality and influence decay of their corresponding items, by studying the relation between the density distributions of}
$\hat{n}^*_v$ and $\hat{\theta}_v$.
For \fakeh{} (\cref{fig:measurement_b}), \revA{we discretize the range of values for $\hat{n}^*$ into $10$ bins, and for each bin we} plot the three-point summaries ($25^{th}$, $75^{th}$ percentiles and median) of $\hat{\theta}$ values of online items \revA{in} \textit{RNIX} and \textit{CNIX}.
\revA{The marginal densities of}
$\hat{n}^*$ and $\hat{\theta}$ are \revA{plotted on the sides of the main panel, and show}
that articles from reputable news sources (\textit{RNIX}) are more \revA{viral} than those from controversial news sources (\textit{CNIX}), while the $\hat{\theta}$ density is similar for \revA{the} two sources. 
\revA{This appears to contradict common intuition, however the joint plot pictures a more nuanced story.
For both \textit{RNIX} and \textit{CNIX}, $\hat{\theta}$ generally decreases as $\hat{n}^*$ increases.
However, for $\hat{n}^* < 0.25$ \textit{CNIX} shows higher values of $\hat{\theta}$, while for $\hat{n}^* > 0.75$ \textit{CNIX} has slower influence decay rates}.
\revA{In other words, low viral controversial articles are forgotten quickly, but highly viral controversial articles are reshared in Twitter for longer than the reputable articles.}

\cref{fig:measurement_a} \revA{shows the marginal} densities of $\hat{n}^*_v$ and $\hat{\theta}_v$ \revA{for} four chosen popular YouTube video categories from \acth{}. 
\revA{The joint plot is less readable than the one in \cref{fig:measurement_b}, and it}
can be found in the online appendix~\citep{appendix}.
We see that \textit{Gaming} videos are substantially less viral than videos from other three categories, but these videos also \revA{show slower influence decay, \rev{indicating} that gamers consume such videos for longer after they were posted.}
We also observe that \textit{News \& Politics} videos \rev{exhibit} similar diffusion patterns as news articles, i.e., \revA{with high virality} and fast decaying influence.

\noindent {\bf \revA{Exploring online item publishers}.} 
\revA{Here, we explore the usage of the diffusion embeddings to analyse the relation between content producers.}
We construct \revA{$\pmb{m}^{n^*}_{\rho}$ the embeddings for a publisher $\rho$ by aggregating the item embedding vectors} ($\pmb{m}^{n^*}_{\rho}$, $\pmb{m}^{c}_{\rho}$ and $\pmb{m}^{\theta}_{\rho}$) \revA{for all online items associated with the $\rho$}.
\revA{Specifically, we compute their element-wise mean} and \revA{we} normalize the vectors to sum to $1$, e.g., $\pmb{m'}^{n^*}_{\rho} = [\sum_{v \in V_{\rho}} \pmb{m}^{n^*}_{v,1}/|V_{\rho}|, \dots]^\mathsf{T}$ and $\pmb{m}^{n^*}_{\rho, i} = \pmb{m'}^{n^*}_{\rho,i} / \sum_{j}\pmb{m'}^{n^*}_{\rho,j}$. 
We compute the distance between two publishers $\rho_1$, $\rho_2$ as $D_{\rho_1, \rho_2}$ following~\cref{eq:embedding_distance}. 
\revA{Finally,} we use t-SNE~\citep{maaten2008visualizing} --- a widely adopted technique for visualizing high dimensional data --- to present \revA{the most} popular publishers in a latent \revA{two-dimensional} space.

\begin{figure}[!tbp]
	\centering
	\begin{subfigure}{0.47\textwidth}
        \includegraphics[width=\textwidth,page=1]{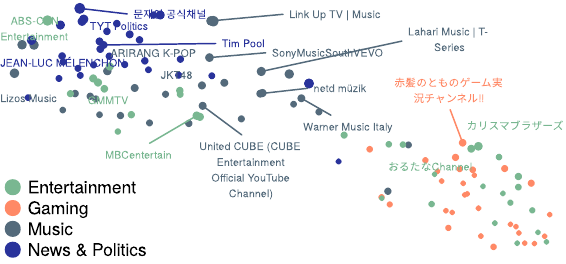}
        \vspace{-4mm}
        \caption{\acth}
        \label{subfig:channel_clustering_a}
    \end{subfigure}
    \begin{subfigure}{0.47\textwidth}
        \includegraphics[width=\textwidth,page=1]{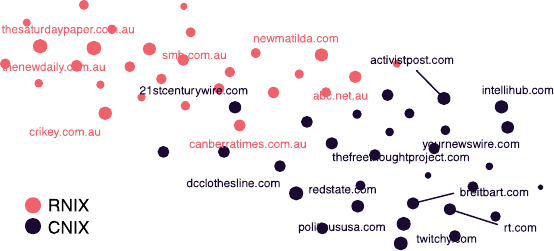}
        \vspace{-4mm}
        \caption{\fakeh}
        \label{subfig:channel_clustering_b}
    \end{subfigure}
    \vspace{-3mm}
	\caption{
        Clustering of publishers with respect to the fitted model parameters. Top $30$ publishers with the most numbers of produced online items in each category are chosen from \acth{} (\textit{Entertainment}, \textit{Music}, \textit{Gaming}, \textit{News \& Politics}) and \fakeh{} (\textit{RNIX}, \textit{CNIX}). Categories of YouTube publishers are determined by their mostly used video categories. The bubble size indicates the average $\hat{n^*}$ of a publisher. Names of $20$ publishers with high average $\hat{n^*}$ values are presented.
	}
    \label{fig:channel_clustering}
    \vspace{-4mm}
\end{figure}

In \cref{fig:channel_clustering}, the top $30$ publishers with the most number of \rev{items of} each category are shown for \acth{} and \fakeh{}.
\revA{For \acth{}, as category is labeled at the item level, we construct publisher categories as the majority category for their items.}
The bubble sizes of individual publishers are scaled by their average $\hat{n}^*$ over all published items.
One conclusion emerges that, in general, publishers from the same category are \revA{also similar} in terms of their diffusion patterns. 
\revA{In both figures}, two major clusters \revA{emerge.
For \acth{} (\cref{subfig:channel_clustering_a}) we observe \textit{Entertainment} and \textit{Gaming} in one cluster, and \textit{Music} and \textit{News \& Politics} in the other.
For \fakeh{} (\cref{subfig:channel_clustering_b}), the \textit{RNIX} and \textit{CNIX} categories appear clearly separable.} 
\cref{subfig:channel_clustering_a} also shows that \textit{Entertainment} is a diverse category
\revA{with its publishers sprinkled across} the entire \revA{latent} space.
\revA{This is due to \textit{Entertainement} videos covering a broad range of subjects, from people singing to online game recordings.} 
\revA{Also}, we identify some viral YouTube publishers such as some K-pop music bands (\textit{ARIRANG K-POP}, \textit{United CUBE}) and controversial news sources including \textit{breitbart.com} from the plots. 
Another interesting observation is that~\cref{subfig:channel_clustering_b} groups similar controversial publishers together, e.g., those showing a strong level of spreading conspiracy (\textit{activistpost.com}, \textit{intellihub.com} and \textit{yournewswire.com}) \revA{in the top-right} corner and those \revA{having} far-right bias in their political stands (\textit{breitbart.com}, \textit{rt.com} and \textit{twitchy.com}) \revA{in the bottom-right corner}.\footnote{Conspiracy levels and political stands can be found in \textit{https://mediabiasfactcheck.com}}

\revA{Given the perceived separability of publisher category in \cref{fig:channel_clustering}, in the next section we setup a predictive exercise.}

\subsection{Prediction of item categories}\label{sec:categorical_prediction}
In this section, we \revA{\rev{build} a predictor for} item category \revA{based on item resharing dynamics and textual content}.
We use three types of features \revA{for online items}: \revA{our proposed diffusion embeddings (see \cref{sec:uncovering})}, temporal features and text features. 
Temporal features are shown \revA{by} previous works to be \revA{useful} in popularity prediction~\citep{bakshy2011everyone,cheng2014can,Mishra2016FeaturePrediction}, but have not been experimented \revA{with} in predicting content categories. 
The text features are \revA{the} natural choices for this task as they carry rich content information, \revA{particularly for the online news articles.} 

\begin{table}[tbp]
	\centering
	\caption{Results of categorical prediction of online items on \acth{} and \fakeh{}. We report Cohen's kappa coefficient and \revA{macro-}F1 of predictors trained with varying combinations of three feature types: the \revA{diffusion embeddings}, temporal features and text features.}
	\vspace{-2mm}
	\begin{tabular}{llrr}
		\toprule
		&  \revA{Features} & Kappa & F1\\
		\midrule
		& \revA{Diffusion Embed.} & 0.289 & 0.488\\
		& Temporal  & 0.536 & 0.675\\
		& \revA{Diffusion Embed.} + Temporal  & 0.540 & 0.679\\
		& Text  & 0.803 & 0.862\\
		& \revA{Diffusion Embed.} + Text  & 0.806 & 0.865\\
		& Text  + Temporal  & 0.830 & 0.883\\
		\multirow{-7}{*}{\raggedright\arraybackslash {\rotatebox[origin=c]{90}{\acth{}}}} & \revA{Diffusion Embed.} + Text  + Temporal  & \textbf{0.831} & \textbf{0.884}\\
		\midrule
		& \revA{Diffusion Embed.}  & 0.610 & 0.675\\
		& Temporal   & 0.840 & 0.872\\
		& \revA{Diffusion Embed.} + Temporal   & 0.844 & 0.874\\
		& Text  & 0.898 & 0.918\\
		& \revA{Diffusion Embed.} + Text  & 0.908 & 0.925\\
		& Text  + Temporal  & 0.930 & 0.944\\
		\multirow{-7}{*}{\raggedright\arraybackslash {\rotatebox[origin=c]{90}{\fakeh{}}}} & \revA{Diffusion Embed.} + Text  + Temporal  & \textbf{0.932} & \textbf{0.945}\\
		\bottomrule
	\end{tabular}
	\label{tab:categorical_prediction}
	\vspace{-2mm}
\end{table}
\begin{figure*}[!htp]
    \centering
    \begin{subfigure}{0.48\textwidth}
        \includegraphics[width=\textwidth]{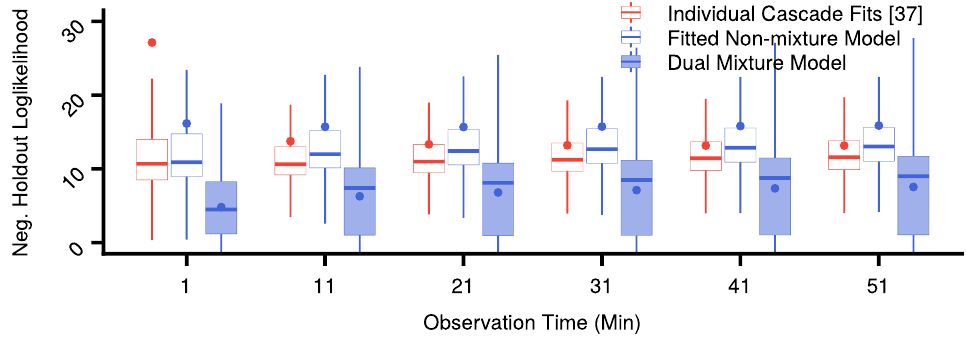}
        \vspace{-5mm}
        \caption{Generalization performance}
        \vspace{-2mm}
        \label{subfig:hll-act}
    \end{subfigure}
    \hspace*{\fill}
	\begin{subfigure}{0.5\textwidth}
		\includegraphics[width=\textwidth,page=1]{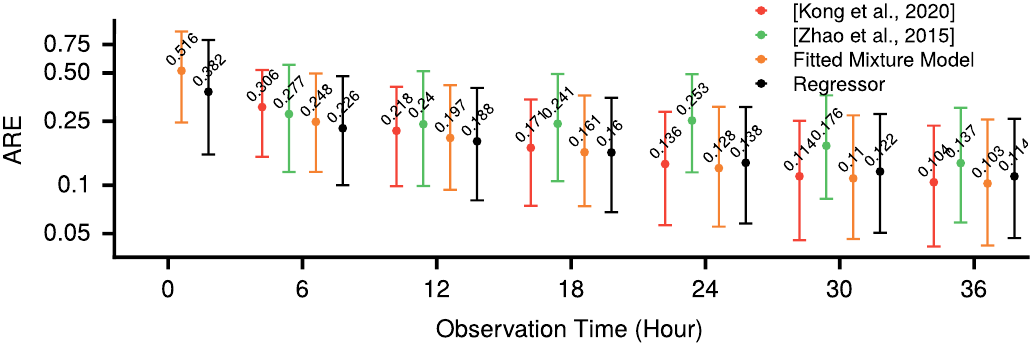}
		\vspace{-5mm}
        \caption{Item popularity prediction}
        \vspace{-2mm}
        \label{subfig:prediction-act}
	\end{subfigure}
	\caption{
        Forecasting for unseen content on \actc{}.
        Fig. (a): \textbf{negative holdout log-likelihood per event} computed from the non-mixture model, the dual mixture models and models from~\citep{Mishra2016FeaturePrediction} given different observation times --- lower is better. The dots are the mean values of the log-likelihoods.
        Fig. (b): item final popularity predictions using four models evaluated with Absolute Relative Error (ARE) --- lower is better. Times at x axis are the observation times since an online item was published. The dots indicate the median values and error bars give the $25^{th}$/$75^{th}$ quantiles of the ARE values.
	}
	\label{fig:hll}
	\vspace{-2mm}
\end{figure*}

\revA{Specifically}, we \revA{build the three types of features as follows.}
\textbf{Diffusion embeddings.} \revA{We use our proposed} diffusion embeddings, i.e., \revA{for a given} online item $v$ \revA{we concatenate the vectors} $\pmb{m}^{n^*}_v$, $\pmb{m}^{c}_v$ and $\pmb{m}^{\theta}_v$.
\textbf{Temporal features.} \revA{We compute the} six-point summaries  (min, mean, median, max, $25^{th}$ and $75^{th}$ percentile) of inter-arrival times, cascade sizes, cascade durations and number of followers of Twitter users involved in cascades.
\textbf{Text features.} 
For each online item, we first concatenate its metadata to a single string. 
This includes video descriptions and video titles for \acth, and article titles, descriptions, keywords and body texts for \fakeh.
We then use the state-of-the-art pre-trained multilingual text embedding model, BERT, to encode text features~\citep{devlin2018bert,Wolf2019HuggingFacesTS}. 
The model encodes at a token-level and generates a set of embeddings with $768$ dimensions for each token. 
We then apply mean aggregations of these embeddings to create a single $768$-dimension vector as the final text feature embedding for the item.

The experiments are conducted on items from the four categories of \acth and all items from \fakeh.
We perform a $50\%$-$50\%$ train-test split and use the \revA{Gradient Boosting Machines} as the predictor (\revA{via the} \texttt{GBM} package in R~\citep{gbm}). 
The predictors' hyper-parameters are selected via $5$-fold cross validations. 
As both datasets are imbalanced, we evaluate the prediction performance \revA{using} the Cohen's kappa coefficient~\citep{viera2005understanding} and the \revA{macro-}F1 score.

\revA{We test all the seven possible combinations of the three feature sets, and}
\cref{tab:categorical_prediction} reports the prediction scores.
When individual \revA{feature set} is \revA{employed}, \revA{the} text features outperform \revA{both} others due to the richer content information \revA{they} contain. 
We \revA{also} note that as generative models are often considered sub-optimal in prediction tasks~\citep{Mishra2016FeaturePrediction}, the \revA{diffusion embeddings appear} the least performing feature types. 
However, \revA{when combined with other feature sets they consistently} provide \revA{a \rev{slight improvement} of performance, indicating that they capture information not present in the textual or temporal features.}
The best predictor is \revA{the one} trained \revA{using} all feature \revA{sets and it} achieves $0.831$ and $0.932$ of kappa values on \acth and \fakeh, respectively. 
The result is particularly interesting on \fakeh, \revA{showing that the temporal features (which are also designed to embed diffusion dynamics) \rev{are} informative features} in predicting controversial news.

\subsection{\revA{Forecasting for unseen content}}
\label{ssec:future-content}
Here, we examine the applications of dual mixture models for modeling unseen diffusion cascades from \actc{}. \rev{The observations are similar on \fakec{} which are discussed in the online appendix~\citep{appendix}.}

\noindent {\bf Generalization performance.} 
\revA{Here}, on individual cascades we compare \revA{the} holdout log-likelihood values of dual mixture models to \revA{single cascade fitting models}~\citep{Mishra2016FeaturePrediction} and to non-mixture \revA{joint} models.
\revA{The dual mixture models and the non-mixture joint models are fitted using cascades from the same publisher.}
Given a cascade $\His_i$ discussing an online item $v$, we find the set of online items produced by the $v$'s publisher, and \revA{we select all cascades relating to} the $5$ most recent items, \revA{denoted as $C_{\rho}$}. 
The dual mixture models are fitted on \revA{all} cascades \revA{in $C_{\rho}$}. 
The holdout log-likelihood is computed via~\cref{eq:expected_hll}. 
\revA{The single cascade models}~\citep{Mishra2016FeaturePrediction} \revA{use power-law Hawkes processes and} fitted on $\His_i(T)$ --- \revA{where $T$ is the maximum time the cascade is observed.}
\revA{Finally, the non-mixture joint} model which is Hawkes processes jointly fitted on all cascades in $C_{\rho}$ \revA{(as discussed in \cref{sec:joint_learning})}.

\cref{subfig:hll-act} \revA{shows the holdout negative log-likelihood} values on \actc{} as boxplots, \revA{trained on increasingly long observation times $T$}. 
\revA{For all} observation times, the dual mixture models consistently outperform \revA{single cascade trained} models and non-mixture models. 
When comparing the \revA{single cascade trained} models \revA{and} the non-mixture \revA{joint} models, despite the former has a better mean and median generalization values, we show in the online appendix~\citep{appendix} that it has more outliers than the latter.
\revA{Finally, we observe that the advantage of using jointly fitted models over single cascade models diminishes as the observations time increases, as the latter observe more data to learn from.}

\noindent {\bf Prediction of final popularity.} 
We compare the final popularity predictions on \actc{} with dual mixture models against \revA{a predictor \rev{built} using} \textit{Seismic}~\citep{Zhao2015SEISMIC:Popularity}, an ensemble model in~\citep{kong2019modeling} and a regressor \revA{trained using} temporal features. 
\textit{Seismic} and the ensemble model predictions are produced by their provided R packages. 
\revA{Since} both models \revA{were designed to predict the} final popularities \revA{of} individual cascades, \revA{we build an item popularity predictor by} following the same steps as in~\cref{sec:prediction} \revA{and using the predictions instead of $\hat{N}_{v,i}(n^*, \Theta^g)$ in \cref{eq:observed-cascade-predicted-size}}.
We construct \revA{the regressor using} the same sets of temporal features as~\cref{sec:categorical_prediction} and
\revA{the tuples (observation times, online items) for the set of examples, and the item final popularity is the dependent variable to predict.
We} train a single regressor using the \texttt{GBM} package in R~\citep{gbm},
\revA{and we obtain predictions for each tuple via} $5$-fold cross validation on \acth{}.
\rev{Finally, }\rev{final popularity predictions of the dual mixture models are computed using~\cref{eq:size_prediction} and at each observation time $T$. We note that we re-fit the BMMs on cascades after the time $T$ in historical cascades to capture changes of content virality in time.}
We evaluate \revA{the} prediction \revA{results using} the Absolute Relative Error (ARE) --- \revA{also} used in~\citep{Zhao2015SEISMIC:Popularity} \revA{and} defined as $\frac{|\hat{N}_v -  N_v|}{N_v}$ where $\hat{N}_v$ and $N_v$ are the predicted popularity and the actual final popularity.

\cref{subfig:prediction-act} summarizes the \revA{prediction} results, \revA{with the} ARE values in log scale. 
As \textit{Seismic} and the ensemble models do not provide cold-start predictions, only results for the dual mixture models and the regressor are presented at $T = 0$ observation time. 
\revA{We see that both the dual mixture models and the temporal features regressor consistently outperform the other two baselines}, \textit{Seismic} and the ensemble model, up to the $18$-hour observation time. 
\revA{Also, the regressor slightly outperforms the dual-mixture model for short observation times}, \revA{after which the dual-mixture model delivers the best predictive performances.}

\section{Conclusion}
This work is concerned with modeling and quantifying temporal dynamics of online items. 
\revA{We start from the observation that maximum likelihood estimates for content virality and influence decay are separable in a Hawkes process, which leads to a separated learning procedure.
Next, we propose a dual mixture self-exciting process, which leverages a Borel mixture model and a kernel mixture model, to jointly model the unfolding of a heterogeneous set of cascades.
When applied to cascades about the same online items, the model directly characterizes the spread dynamics of online items and supplies interpretable quantities as well as methods for predicting the final content popularities}.

\noindent {\bf Limitations and future work.} 
Due to the restriction of the size distribution of Hawkes processes, the current joint fitting is restricted to complete and unmarked processes.
We plan to relax these constraints to allow for joint modeling with more flexible forms of Hawkes processes.

\noindent\textbf{\small{Acknowledgments.}}
\small{This research is supported by Facebook Research under the content policy grants, the Asian Office of Aerospace Research and Development (AOARD) Grant 19IOA078, Australian Research Council Discovery Project DP180101985 and the Data61, CSIRO PhD scholarship. We also thank the National Computational Infrastructure (NCI) for providing computational resources, supported by the Australian Government. Access to the \textit{RNCNIX} dataset was provided courtesy of the Digital Media Research Centre at Queensland University of Technology, and is supported by the Australian Research Council projects FT130100703 and DP200101317.
}

\bibliographystyle{ACM-Reference-Format}
\bibliography{acm}


\begin{thebibliography}{2}


\ifx \showCODEN    \undefined \def \showCODEN     #1{\unskip}     \fi
\ifx \showDOI      \undefined \def \showDOI       #1{#1}\fi
\ifx \showISBNx    \undefined \def \showISBNx     #1{\unskip}     \fi
\ifx \showISBNxiii \undefined \def \showISBNxiii  #1{\unskip}     \fi
\ifx \showISSN     \undefined \def \showISSN      #1{\unskip}     \fi
\ifx \showLCCN     \undefined \def \showLCCN      #1{\unskip}     \fi
\ifx \shownote     \undefined \def \shownote      #1{#1}          \fi
\ifx \showarticletitle \undefined \def \showarticletitle #1{#1}   \fi
\ifx \showURL      \undefined \def \showURL       {\relax}        \fi
\providecommand\bibfield[2]{#2}
\providecommand\bibinfo[2]{#2}
\providecommand\natexlab[1]{#1}
\providecommand\showeprint[2][]{arXiv:#2}

\bibitem[\protect\citeauthoryear{Maaten and Hinton}{Maaten and Hinton}{2008}]%
        {maaten2008visualizing}
\bibfield{author}{\bibinfo{person}{Laurens van~der Maaten} {and}
  \bibinfo{person}{Geoffrey Hinton}.} \bibinfo{year}{2008}\natexlab{}.
\newblock \showarticletitle{Visualizing data using t-SNE}.
\newblock \bibinfo{journal}{\emph{JMLR}} (\bibinfo{year}{2008}).
\newblock


\bibitem[\protect\citeauthoryear{Tomasi}{Tomasi}{2004}]%
        {tomasi2004estimating}
\bibfield{author}{\bibinfo{person}{Carlo Tomasi}.}
  \bibinfo{year}{2004}\natexlab{}.
\newblock \showarticletitle{Estimating Gaussian mixture densities with em--a
  tutorial}.
\newblock \bibinfo{journal}{\emph{Duke University}} (\bibinfo{year}{2004}).
\newblock


\end{thebibliography}



\begin{thebibliography}{54}


\ifx \showCODEN    \undefined \def \showCODEN     #1{\unskip}     \fi
\ifx \showDOI      \undefined \def \showDOI       #1{#1}\fi
\ifx \showISBNx    \undefined \def \showISBNx     #1{\unskip}     \fi
\ifx \showISBNxiii \undefined \def \showISBNxiii  #1{\unskip}     \fi
\ifx \showISSN     \undefined \def \showISSN      #1{\unskip}     \fi
\ifx \showLCCN     \undefined \def \showLCCN      #1{\unskip}     \fi
\ifx \shownote     \undefined \def \shownote      #1{#1}          \fi
\ifx \showarticletitle \undefined \def \showarticletitle #1{#1}   \fi
\ifx \showURL      \undefined \def \showURL       {\relax}        \fi
\providecommand\bibfield[2]{#2}
\providecommand\bibinfo[2]{#2}
\providecommand\natexlab[1]{#1}
\providecommand\showeprint[2][]{arXiv:#2}

\bibitem[\protect\citeauthoryear{Apostolopoulou, Linderman, Miller, and
  Dubrawski}{Apostolopoulou et~al\mbox{.}}{2019}]%
        {apostolopoulou2019mutually}
\bibfield{author}{\bibinfo{person}{Ifigeneia Apostolopoulou},
  \bibinfo{person}{Scott Linderman}, \bibinfo{person}{Kyle Miller}, {and}
  \bibinfo{person}{Artur Dubrawski}.} \bibinfo{year}{2019}\natexlab{}.
\newblock \showarticletitle{Mutually Regressive Point Processes}. In
  \bibinfo{booktitle}{\emph{NeurIPS}}.
\newblock


\bibitem[\protect\citeauthoryear{Appendix}{Appendix}{2020}]%
        {appendix}
\bibfield{author}{\bibinfo{person}{Appendix}.} \bibinfo{year}{2020}\natexlab{}.
\newblock \bibinfo{title}{Appendix: \titlename}.
\newblock
\newblock
\newblock
\shownote{\url{https://bit.ly/3kDZbvq}.}


\bibitem[\protect\citeauthoryear{Arjovsky, Chintala, and Bottou}{Arjovsky
  et~al\mbox{.}}{2017}]%
        {arjovsky2017wasserstein}
\bibfield{author}{\bibinfo{person}{Martin Arjovsky}, \bibinfo{person}{Soumith
  Chintala}, {and} \bibinfo{person}{L{\'e}on Bottou}.}
  \bibinfo{year}{2017}\natexlab{}.
\newblock \showarticletitle{Wasserstein gan}.
\newblock \bibinfo{journal}{\emph{arXiv}} (\bibinfo{year}{2017}).
\newblock


\bibitem[\protect\citeauthoryear{Bacry, Mastromatteo, and Muzy}{Bacry
  et~al\mbox{.}}{2015}]%
        {bacry2015hawkes}
\bibfield{author}{\bibinfo{person}{Emmanuel Bacry}, \bibinfo{person}{Iacopo
  Mastromatteo}, {and} \bibinfo{person}{Jean-Fran{\c{c}}ois Muzy}.}
  \bibinfo{year}{2015}\natexlab{}.
\newblock \showarticletitle{Hawkes processes in finance}.
\newblock \bibinfo{journal}{\emph{Market Microstructure and Liquidity}}
  (\bibinfo{year}{2015}).
\newblock


\bibitem[\protect\citeauthoryear{Bakshy, Hofman, Mason, and Watts}{Bakshy
  et~al\mbox{.}}{2011}]%
        {bakshy2011everyone}
\bibfield{author}{\bibinfo{person}{Eytan Bakshy}, \bibinfo{person}{Jake~M
  Hofman}, \bibinfo{person}{Winter~A Mason}, {and} \bibinfo{person}{Duncan~J
  Watts}.} \bibinfo{year}{2011}\natexlab{}.
\newblock \showarticletitle{Everyone's an influencer: quantifying influence on
  twitter}. In \bibinfo{booktitle}{\emph{WSDM}}.
\newblock


\bibitem[\protect\citeauthoryear{Bao}{Bao}{2016}]%
        {bao2016modeling}
\bibfield{author}{\bibinfo{person}{Peng Bao}.} \bibinfo{year}{2016}\natexlab{}.
\newblock \showarticletitle{Modeling and predicting popularity dynamics via an
  influence-based self-excited Hawkes process}. In
  \bibinfo{booktitle}{\emph{CIKM}}.
\newblock


\bibitem[\protect\citeauthoryear{Bishop}{Bishop}{2006}]%
        {bishop2006pattern}
\bibfield{author}{\bibinfo{person}{Christopher~M Bishop}.}
  \bibinfo{year}{2006}\natexlab{}.
\newblock \bibinfo{booktitle}{\emph{Pattern recognition and machine learning}}.
\newblock \bibinfo{publisher}{springer}.
\newblock


\bibitem[\protect\citeauthoryear{Borel}{Borel}{1942}]%
        {borel1942emploi}
\bibfield{author}{\bibinfo{person}{{\'E}mile Borel}.}
  \bibinfo{year}{1942}\natexlab{}.
\newblock \showarticletitle{Sur l'emploi du th{\'e}oreme de Bernoulli pour
  faciliter le calcul d'une infinit{\'e} de coefficients. Application au
  probleme de l'attentea un guichet}.
\newblock \bibinfo{journal}{\emph{CR Acad. Sci. Paris}} (\bibinfo{year}{1942}).
\newblock


\bibitem[\protect\citeauthoryear{Bruns}{Bruns}{2016}]%
        {bruns2016big}
\bibfield{author}{\bibinfo{person}{Axel Bruns}.}
  \bibinfo{year}{2016}\natexlab{}.
\newblock \showarticletitle{Big Data Analysis}.
\newblock \bibinfo{journal}{\emph{The Sage handbook of digital journalism}}
  (\bibinfo{year}{2016}).
\newblock


\bibitem[\protect\citeauthoryear{Bruns}{Bruns}{2017}]%
        {bruns2016publics}
\bibfield{author}{\bibinfo{person}{Axel Bruns}.}
  \bibinfo{year}{2017}\natexlab{}.
\newblock \showarticletitle{Making Audience Engagement Visible: Publics for
  Journalism on Social Media Platforms}.
\newblock \bibinfo{journal}{\emph{The Routledge Companion to Digital Journalism
  Studies}} (\bibinfo{year}{2017}).
\newblock


\bibitem[\protect\citeauthoryear{Bruns and Keller}{Bruns and Keller}{2020}]%
        {bruns2020news}
\bibfield{author}{\bibinfo{person}{Axel Bruns} {and} \bibinfo{person}{Tobias
  Keller}.} \bibinfo{year}{2020}\natexlab{}.
\newblock \showarticletitle{News diffusion on Twitter: Comparing the
  dissemination careers for mainstream and marginal news}. In
  \bibinfo{booktitle}{\emph{Social Media \& Society 2020 Conference}}.
\newblock


\bibitem[\protect\citeauthoryear{Burnham and Anderson}{Burnham and
  Anderson}{2004}]%
        {burnham2004multimodel}
\bibfield{author}{\bibinfo{person}{Kenneth~P Burnham} {and}
  \bibinfo{person}{David~R Anderson}.} \bibinfo{year}{2004}\natexlab{}.
\newblock \showarticletitle{Multimodel inference: understanding AIC and BIC in
  model selection}.
\newblock \bibinfo{journal}{\emph{Sociological methods \& research}}
  (\bibinfo{year}{2004}).
\newblock


\bibitem[\protect\citeauthoryear{Cao, Shen, Cen, Ouyang, and Cheng}{Cao
  et~al\mbox{.}}{2017}]%
        {cao2017deephawkes}
\bibfield{author}{\bibinfo{person}{Qi Cao}, \bibinfo{person}{Huawei Shen},
  \bibinfo{person}{Keting Cen}, \bibinfo{person}{Wentao Ouyang}, {and}
  \bibinfo{person}{Xueqi Cheng}.} \bibinfo{year}{2017}\natexlab{}.
\newblock \showarticletitle{Deephawkes: Bridging the gap between prediction and
  understanding of information cascades}. In \bibinfo{booktitle}{\emph{CIKM}}.
\newblock


\bibitem[\protect\citeauthoryear{Cheng, Adamic, Dow, Kleinberg, and
  Leskovec}{Cheng et~al\mbox{.}}{2014}]%
        {cheng2014can}
\bibfield{author}{\bibinfo{person}{Justin Cheng}, \bibinfo{person}{Lada
  Adamic}, \bibinfo{person}{P~Alex Dow}, \bibinfo{person}{Jon~Michael
  Kleinberg}, {and} \bibinfo{person}{Jure Leskovec}.}
  \bibinfo{year}{2014}\natexlab{}.
\newblock \showarticletitle{Can cascades be predicted?}. In
  \bibinfo{booktitle}{\emph{WWW}}.
\newblock


\bibitem[\protect\citeauthoryear{Daley and Vere-Jones}{Daley and
  Vere-Jones}{2008}]%
        {Daley2008}
\bibfield{author}{\bibinfo{person}{Daryl~J Daley} {and} \bibinfo{person}{David
  Vere-Jones}.} \bibinfo{year}{2008}\natexlab{}.
\newblock \showarticletitle{Conditional Intensities and Likelihoods}.
\newblock In \bibinfo{booktitle}{\emph{An introduction to the theory of point
  processes}}. Vol.~\bibinfo{volume}{I}. \bibinfo{publisher}{Springer}, Chapter
  7.2.
\newblock
\showISBNx{978-0-387-21337-8}


\bibitem[\protect\citeauthoryear{Daw and Pender}{Daw and Pender}{2018}]%
        {daw2018queue}
\bibfield{author}{\bibinfo{person}{Andrew Daw} {and} \bibinfo{person}{Jamol
  Pender}.} \bibinfo{year}{2018}\natexlab{}.
\newblock \showarticletitle{The Queue-Hawkes Process: Ephemeral
  Self-Excitement}.
\newblock \bibinfo{journal}{\emph{arXiv}} (\bibinfo{year}{2018}).
\newblock


\bibitem[\protect\citeauthoryear{Dempster, Laird, and Rubin}{Dempster
  et~al\mbox{.}}{1977}]%
        {dempster1977maximum}
\bibfield{author}{\bibinfo{person}{Arthur~P Dempster}, \bibinfo{person}{Nan~M
  Laird}, {and} \bibinfo{person}{Donald~B Rubin}.}
  \bibinfo{year}{1977}\natexlab{}.
\newblock \showarticletitle{Maximum likelihood from incomplete data via the EM
  algorithm}.
\newblock \bibinfo{journal}{\emph{Journal of the Royal Statistical Society:
  Series B (Methodological)}} (\bibinfo{year}{1977}).
\newblock


\bibitem[\protect\citeauthoryear{Devlin, Chang, Lee, and Toutanova}{Devlin
  et~al\mbox{.}}{2018}]%
        {devlin2018bert}
\bibfield{author}{\bibinfo{person}{Jacob Devlin}, \bibinfo{person}{Ming-Wei
  Chang}, \bibinfo{person}{Kenton Lee}, {and} \bibinfo{person}{Kristina
  Toutanova}.} \bibinfo{year}{2018}\natexlab{}.
\newblock \showarticletitle{Bert: Pre-training of deep bidirectional
  transformers for language understanding}.
\newblock \bibinfo{journal}{\emph{arXiv}} (\bibinfo{year}{2018}).
\newblock


\bibitem[\protect\citeauthoryear{Du, Farajtabar, Ahmed, Smola, and Song}{Du
  et~al\mbox{.}}{2015}]%
        {du2015dirichlet}
\bibfield{author}{\bibinfo{person}{Nan Du}, \bibinfo{person}{Mehrdad
  Farajtabar}, \bibinfo{person}{Amr Ahmed}, \bibinfo{person}{Alexander~J
  Smola}, {and} \bibinfo{person}{Le Song}.} \bibinfo{year}{2015}\natexlab{}.
\newblock \showarticletitle{Dirichlet-hawkes processes with applications to
  clustering continuous-time document streams}. In
  \bibinfo{booktitle}{\emph{KDD}}. ACM.
\newblock


\bibitem[\protect\citeauthoryear{Durrett}{Durrett}{2010}]%
        {durrett2019probability}
\bibfield{author}{\bibinfo{person}{Rick Durrett}.}
  \bibinfo{year}{2010}\natexlab{}.
\newblock \bibinfo{booktitle}{\emph{Probability: theory and examples}}.
\newblock \bibinfo{publisher}{Cambridge university press}.
\newblock


\bibitem[\protect\citeauthoryear{Goel, Watts, and Goldstein}{Goel
  et~al\mbox{.}}{2012}]%
        {goel2012structure}
\bibfield{author}{\bibinfo{person}{Sharad Goel}, \bibinfo{person}{Duncan~J
  Watts}, {and} \bibinfo{person}{Daniel~G Goldstein}.}
  \bibinfo{year}{2012}\natexlab{}.
\newblock \showarticletitle{The structure of online diffusion networks}. In
  \bibinfo{booktitle}{\emph{Proceedings of the 13th ACM Conference on
  Electronic Commerce}}.
\newblock


\bibitem[\protect\citeauthoryear{Gomez-Rodriguez, Balduzzi, and
  Sch{\"o}lkopf}{Gomez-Rodriguez et~al\mbox{.}}{2011}]%
        {rodriguez2011uncovering}
\bibfield{author}{\bibinfo{person}{Manuel Gomez-Rodriguez},
  \bibinfo{person}{David Balduzzi}, {and} \bibinfo{person}{Bernhard
  Sch{\"o}lkopf}.} \bibinfo{year}{2011}\natexlab{}.
\newblock \showarticletitle{Uncovering the temporal dynamics of diffusion
  networks}. In \bibinfo{booktitle}{\emph{ICML}}.
\newblock


\bibitem[\protect\citeauthoryear{Greenwell, Boehmke, Cunningham, and
  Developers}{Greenwell et~al\mbox{.}}{2019}]%
        {gbm}
\bibfield{author}{\bibinfo{person}{Brandon Greenwell}, \bibinfo{person}{Bradley
  Boehmke}, \bibinfo{person}{Jay Cunningham}, {and} \bibinfo{person}{GBM
  Developers}.} \bibinfo{year}{2019}\natexlab{}.
\newblock \bibinfo{booktitle}{\emph{gbm: Generalized Boosted Regression
  Models}}.
\newblock
\newblock
\shownote{R package version 2.1.5.}


\bibitem[\protect\citeauthoryear{Haight and Breuer}{Haight and Breuer}{1960}]%
        {Haight1960}
\bibfield{author}{\bibinfo{person}{Frank~A. Haight} {and}
  \bibinfo{person}{Melvin~Allen Breuer}.} \bibinfo{year}{1960}\natexlab{}.
\newblock \showarticletitle{{The Borel-Tanner Distribution}}.
\newblock \bibinfo{journal}{\emph{Biometrika}} (\bibinfo{year}{1960}).
\newblock


\bibitem[\protect\citeauthoryear{Harrell~Jr, Dupont, et~al\mbox{.}}{Harrell~Jr
  et~al\mbox{.}}{2017}]%
        {harrell2017hmisc}
\bibfield{author}{\bibinfo{person}{Frank~E Harrell~Jr},
  \bibinfo{person}{Charles Dupont}, {et~al\mbox{.}}}
  \bibinfo{year}{2017}\natexlab{}.
\newblock \showarticletitle{Hmisc: Harrell miscellaneous. R package version
  4.0-3}.
\newblock \bibinfo{journal}{\emph{Online publication}} (\bibinfo{year}{2017}).
\newblock


\bibitem[\protect\citeauthoryear{Hawkes}{Hawkes}{1971}]%
        {hawkes1971spectra}
\bibfield{author}{\bibinfo{person}{Alan~G Hawkes}.}
  \bibinfo{year}{1971}\natexlab{}.
\newblock \showarticletitle{Spectra of some self-exciting and mutually exciting
  point processes}.
\newblock \bibinfo{journal}{\emph{Biometrika}} (\bibinfo{year}{1971}).
\newblock


\bibitem[\protect\citeauthoryear{Hawkes and Oakes}{Hawkes and Oakes}{1974}]%
        {hawkes1974cluster}
\bibfield{author}{\bibinfo{person}{Alan~G Hawkes} {and} \bibinfo{person}{David
  Oakes}.} \bibinfo{year}{1974}\natexlab{}.
\newblock \showarticletitle{A cluster process representation of a self-exciting
  process}.
\newblock \bibinfo{journal}{\emph{Journal of Applied Probability}}
  (\bibinfo{year}{1974}).
\newblock


\bibitem[\protect\citeauthoryear{Kobayashi and Lambiotte}{Kobayashi and
  Lambiotte}{2016}]%
        {Kobayashi2016TiDeH:Dynamics}
\bibfield{author}{\bibinfo{person}{Ryota Kobayashi} {and}
  \bibinfo{person}{Renaud Lambiotte}.} \bibinfo{year}{2016}\natexlab{}.
\newblock \showarticletitle{TiDeH: Time-Dependent Hawkes Process for Predicting
  Retweet Dynamics}. In \bibinfo{booktitle}{\emph{ICWSM}}.
\newblock


\bibitem[\protect\citeauthoryear{Kong, Rizoiu, and Xie}{Kong
  et~al\mbox{.}}{2020}]%
        {kong2019modeling}
\bibfield{author}{\bibinfo{person}{Quyu Kong}, \bibinfo{person}{Marian-Andrei
  Rizoiu}, {and} \bibinfo{person}{Lexing Xie}.}
  \bibinfo{year}{2020}\natexlab{}.
\newblock \showarticletitle{Modeling Information Cascades with Self-exciting
  Processes via Generalized Epidemic Models}. In
  \bibinfo{booktitle}{\emph{WSDM}}.
\newblock


\bibitem[\protect\citeauthoryear{Lakkaraju, McAuley, and Leskovec}{Lakkaraju
  et~al\mbox{.}}{2013}]%
        {lakkaraju2013s}
\bibfield{author}{\bibinfo{person}{Himabindu Lakkaraju},
  \bibinfo{person}{Julian McAuley}, {and} \bibinfo{person}{Jure Leskovec}.}
  \bibinfo{year}{2013}\natexlab{}.
\newblock \showarticletitle{What's in a name? understanding the interplay
  between titles, content, and communities in social media}. In
  \bibinfo{booktitle}{\emph{ICWSM}}.
\newblock


\bibitem[\protect\citeauthoryear{Laub, Taimre, and Pollett}{Laub
  et~al\mbox{.}}{2015}]%
        {laub2015hawkes}
\bibfield{author}{\bibinfo{person}{Patrick~J Laub}, \bibinfo{person}{Thomas
  Taimre}, {and} \bibinfo{person}{Philip~K Pollett}.}
  \bibinfo{year}{2015}\natexlab{}.
\newblock \showarticletitle{Hawkes processes}.
\newblock \bibinfo{journal}{\emph{arXiv}} (\bibinfo{year}{2015}).
\newblock


\bibitem[\protect\citeauthoryear{Luko{\v{c}}ien{\.e} and
  Vermunt}{Luko{\v{c}}ien{\.e} and Vermunt}{2009}]%
        {lukovciene2009determining}
\bibfield{author}{\bibinfo{person}{Olga Luko{\v{c}}ien{\.e}} {and}
  \bibinfo{person}{Jeroen~K Vermunt}.} \bibinfo{year}{2009}\natexlab{}.
\newblock \showarticletitle{Determining the number of components in mixture
  models for hierarchical data}.
\newblock In \bibinfo{booktitle}{\emph{Advances in data analysis, data handling
  and business intelligence}}. \bibinfo{publisher}{Springer}.
\newblock


\bibitem[\protect\citeauthoryear{Ma, Gao, Mitra, Kwon, Jansen, Wong, and
  Cha}{Ma et~al\mbox{.}}{2016}]%
        {ma2016detecting}
\bibfield{author}{\bibinfo{person}{Jing Ma}, \bibinfo{person}{Wei Gao},
  \bibinfo{person}{Prasenjit Mitra}, \bibinfo{person}{Sejeong Kwon},
  \bibinfo{person}{Bernard~J Jansen}, \bibinfo{person}{Kam-Fai Wong}, {and}
  \bibinfo{person}{Meeyoung Cha}.} \bibinfo{year}{2016}\natexlab{}.
\newblock \showarticletitle{Detecting rumors from microblogs with recurrent
  neural networks.}. In \bibinfo{booktitle}{\emph{IJCAI}}.
\newblock


\bibitem[\protect\citeauthoryear{Maaten and Hinton}{Maaten and Hinton}{2008}]%
        {maaten2008visualizing}
\bibfield{author}{\bibinfo{person}{Laurens van~der Maaten} {and}
  \bibinfo{person}{Geoffrey Hinton}.} \bibinfo{year}{2008}\natexlab{}.
\newblock \showarticletitle{Visualizing data using t-SNE}.
\newblock \bibinfo{journal}{\emph{JMLR}} (\bibinfo{year}{2008}).
\newblock


\bibitem[\protect\citeauthoryear{Martin, Hofman, Sharma, Anderson, and
  Watts}{Martin et~al\mbox{.}}{2016}]%
        {martin2016exploring}
\bibfield{author}{\bibinfo{person}{Travis Martin}, \bibinfo{person}{Jake~M
  Hofman}, \bibinfo{person}{Amit Sharma}, \bibinfo{person}{Ashton Anderson},
  {and} \bibinfo{person}{Duncan~J Watts}.} \bibinfo{year}{2016}\natexlab{}.
\newblock \showarticletitle{Exploring limits to prediction in complex social
  systems}. In \bibinfo{booktitle}{\emph{WWW}}.
\newblock


\bibitem[\protect\citeauthoryear{Mishra}{Mishra}{2019}]%
        {mishra2019linking}
\bibfield{author}{\bibinfo{person}{Swapnil Mishra}.}
  \bibinfo{year}{2019}\natexlab{}.
\newblock \emph{\bibinfo{title}{Linking Models for Collective Attention in
  Social Media}}.
\newblock \bibinfo{thesistype}{Ph.D. Dissertation}.
\newblock


\bibitem[\protect\citeauthoryear{Mishra, Rizoiu, and Xie}{Mishra
  et~al\mbox{.}}{2016}]%
        {Mishra2016FeaturePrediction}
\bibfield{author}{\bibinfo{person}{Swapnil Mishra},
  \bibinfo{person}{Marian-Andrei Rizoiu}, {and} \bibinfo{person}{Lexing Xie}.}
  \bibinfo{year}{2016}\natexlab{}.
\newblock \showarticletitle{Feature Driven and Point Process Approaches for
  Popularity Prediction}. In \bibinfo{booktitle}{\emph{CIKM}}.
\newblock


\bibitem[\protect\citeauthoryear{O'Brien, Aleta, Moreno, and Gleeson}{O'Brien
  et~al\mbox{.}}{2020}]%
        {o2020quantifying}
\bibfield{author}{\bibinfo{person}{Joseph~D O'Brien}, \bibinfo{person}{Alberto
  Aleta}, \bibinfo{person}{Yamir Moreno}, {and} \bibinfo{person}{James~P
  Gleeson}.} \bibinfo{year}{2020}\natexlab{}.
\newblock \showarticletitle{Quantifying Uncertainty in a Predictive Model for
  Popularity Dynamics}.
\newblock \bibinfo{journal}{\emph{arXiv}} (\bibinfo{year}{2020}).
\newblock


\bibitem[\protect\citeauthoryear{Ogata}{Ogata}{1988}]%
        {ogata1988statistical}
\bibfield{author}{\bibinfo{person}{Yosihiko Ogata}.}
  \bibinfo{year}{1988}\natexlab{}.
\newblock \showarticletitle{Statistical models for earthquake occurrences and
  residual analysis for point processes}.
\newblock \bibinfo{journal}{\emph{J. Amer. Statist. Assoc.}}
  (\bibinfo{year}{1988}).
\newblock


\bibitem[\protect\citeauthoryear{Parmar, Bushi, Bhattacharya, and Kumar}{Parmar
  et~al\mbox{.}}{2017}]%
        {parmar2017forecasting}
\bibfield{author}{\bibinfo{person}{Krunal Parmar}, \bibinfo{person}{Samuel
  Bushi}, \bibinfo{person}{Sourangshu Bhattacharya}, {and}
  \bibinfo{person}{Surender Kumar}.} \bibinfo{year}{2017}\natexlab{}.
\newblock \showarticletitle{Forecasting ad-impressions on online retail
  websites using non-homogeneous hawkes processes}. In
  \bibinfo{booktitle}{\emph{CIKM}}.
\newblock


\bibitem[\protect\citeauthoryear{Rizoiu, Mishra, Kong, Carman, and Xie}{Rizoiu
  et~al\mbox{.}}{2018}]%
        {Rizoiu2017c}
\bibfield{author}{\bibinfo{person}{Marian-Andrei Rizoiu},
  \bibinfo{person}{Swapnil Mishra}, \bibinfo{person}{Quyu Kong},
  \bibinfo{person}{Mark Carman}, {and} \bibinfo{person}{Lexing Xie}.}
  \bibinfo{year}{2018}\natexlab{}.
\newblock \showarticletitle{SIR-Hawkes: on the Relationship Between Epidemic
  Models and Hawkes Point Processes}. In \bibinfo{booktitle}{\emph{WWW}}.
\newblock


\bibitem[\protect\citeauthoryear{Rizoiu, Xie, Sanner, Cebrian, Yu, and {Van
  Hentenryck}}{Rizoiu et~al\mbox{.}}{2017}]%
        {Rizoiu2016ExpectingPopularity}
\bibfield{author}{\bibinfo{person}{Marian-Andrei Rizoiu},
  \bibinfo{person}{Lexing Xie}, \bibinfo{person}{Scott Sanner},
  \bibinfo{person}{Manuel Cebrian}, \bibinfo{person}{Honglin Yu}, {and}
  \bibinfo{person}{Pascal {Van Hentenryck}}.} \bibinfo{year}{2017}\natexlab{}.
\newblock \showarticletitle{Expecting to be HIP: Hawkes Intensity Processes for
  Social Media Popularity}. In \bibinfo{booktitle}{\emph{WWW}}.
\newblock


\bibitem[\protect\citeauthoryear{Tan, Lee, and Pang}{Tan et~al\mbox{.}}{2014}]%
        {tan2014effect}
\bibfield{author}{\bibinfo{person}{Chenhao Tan}, \bibinfo{person}{Lillian Lee},
  {and} \bibinfo{person}{Bo Pang}.} \bibinfo{year}{2014}\natexlab{}.
\newblock \showarticletitle{The effect of wording on message propagation:
  Topic-and author-controlled natural experiments on Twitter}. In
  \bibinfo{booktitle}{\emph{ACL}}.
\newblock


\bibitem[\protect\citeauthoryear{Tomasi}{Tomasi}{2004}]%
        {tomasi2004estimating}
\bibfield{author}{\bibinfo{person}{Carlo Tomasi}.}
  \bibinfo{year}{2004}\natexlab{}.
\newblock \showarticletitle{Estimating Gaussian mixture densities with em--a
  tutorial}.
\newblock \bibinfo{journal}{\emph{Duke University}} (\bibinfo{year}{2004}).
\newblock


\bibitem[\protect\citeauthoryear{Viera, Garrett, et~al\mbox{.}}{Viera
  et~al\mbox{.}}{2005}]%
        {viera2005understanding}
\bibfield{author}{\bibinfo{person}{Anthony~J Viera}, \bibinfo{person}{Joanne~M
  Garrett}, {et~al\mbox{.}}} \bibinfo{year}{2005}\natexlab{}.
\newblock \showarticletitle{Understanding interobserver agreement: the kappa
  statistic}.
\newblock \bibinfo{journal}{\emph{Fam med}} (\bibinfo{year}{2005}).
\newblock


\bibitem[\protect\citeauthoryear{W{\"{a}}chter and Biegler}{W{\"{a}}chter and
  Biegler}{2006}]%
        {Wachter2006}
\bibfield{author}{\bibinfo{person}{A W{\"{a}}chter} {and} \bibinfo{person}{L~T
  Biegler}.} \bibinfo{year}{2006}\natexlab{}.
\newblock \showarticletitle{{On the Implementation of a Primal-Dual Interior
  Point Filter Line Search Algorithm for Large-Scale Nonlinear Programming}}.
\newblock \bibinfo{journal}{\emph{Mathematical Programming}}
  (\bibinfo{year}{2006}).
\newblock


\bibitem[\protect\citeauthoryear{Wolf, Debut, Sanh, Chaumond, Delangue, Moi,
  Cistac, Rault, Louf, Funtowicz, and Brew}{Wolf et~al\mbox{.}}{2019}]%
        {Wolf2019HuggingFacesTS}
\bibfield{author}{\bibinfo{person}{Thomas Wolf}, \bibinfo{person}{Lysandre
  Debut}, \bibinfo{person}{Victor Sanh}, \bibinfo{person}{Julien Chaumond},
  \bibinfo{person}{Clement Delangue}, \bibinfo{person}{Anthony Moi},
  \bibinfo{person}{Pierric Cistac}, \bibinfo{person}{Tim Rault},
  \bibinfo{person}{R'emi Louf}, \bibinfo{person}{Morgan Funtowicz}, {and}
  \bibinfo{person}{Jamie Brew}.} \bibinfo{year}{2019}\natexlab{}.
\newblock \showarticletitle{HuggingFace's Transformers: State-of-the-art
  Natural Language Processing}.
\newblock \bibinfo{journal}{\emph{arXiv}} (\bibinfo{year}{2019}).
\newblock


\bibitem[\protect\citeauthoryear{Wu, Rizoiu, and Xie}{Wu et~al\mbox{.}}{2018}]%
        {wu2018beyond}
\bibfield{author}{\bibinfo{person}{Siqi Wu}, \bibinfo{person}{Marian-Andrei
  Rizoiu}, {and} \bibinfo{person}{Lexing Xie}.}
  \bibinfo{year}{2018}\natexlab{}.
\newblock \showarticletitle{Beyond views: Measuring and predicting engagement
  in online videos}. In \bibinfo{booktitle}{\emph{ICWSM}}.
\newblock


\bibitem[\protect\citeauthoryear{Wu, Yan, Yang, and Zha}{Wu
  et~al\mbox{.}}{2020}]%
        {Wu2020discovering}
\bibfield{author}{\bibinfo{person}{Weichang Wu}, \bibinfo{person}{Junchi Yan},
  \bibinfo{person}{Xiaokang Yang}, {and} \bibinfo{person}{Hongyuan Zha}.}
  \bibinfo{year}{2020}\natexlab{}.
\newblock \showarticletitle{{Discovering Temporal Patterns for Event Sequence
  Clustering via Policy Mixture Model}}.
\newblock \bibinfo{journal}{\emph{IEEE Transactions on Knowledge and Data
  Engineering}} (\bibinfo{year}{2020}).
\newblock


\bibitem[\protect\citeauthoryear{Xu, Farajtabar, and Zha}{Xu
  et~al\mbox{.}}{2016}]%
        {xu2016learning}
\bibfield{author}{\bibinfo{person}{Hongteng Xu}, \bibinfo{person}{Mehrdad
  Farajtabar}, {and} \bibinfo{person}{Hongyuan Zha}.}
  \bibinfo{year}{2016}\natexlab{}.
\newblock \showarticletitle{Learning granger causality for hawkes processes}.
  In \bibinfo{booktitle}{\emph{ICML}}.
\newblock


\bibitem[\protect\citeauthoryear{Xu and Zha}{Xu and Zha}{2017}]%
        {xu2017dirichlet}
\bibfield{author}{\bibinfo{person}{Hongteng Xu} {and} \bibinfo{person}{Hongyuan
  Zha}.} \bibinfo{year}{2017}\natexlab{}.
\newblock \showarticletitle{A Dirichlet mixture model of Hawkes processes for
  event sequence clustering}. In \bibinfo{booktitle}{\emph{NeurIPS}}.
\newblock


\bibitem[\protect\citeauthoryear{Yang and Zha}{Yang and Zha}{2013}]%
        {yang2013mixture}
\bibfield{author}{\bibinfo{person}{Shuang-Hong Yang} {and}
  \bibinfo{person}{Hongyuan Zha}.} \bibinfo{year}{2013}\natexlab{}.
\newblock \showarticletitle{Mixture of mutually exciting processes for viral
  diffusion}. In \bibinfo{booktitle}{\emph{ICML}}.
\newblock


\bibitem[\protect\citeauthoryear{Zhang, Walder, Rizoiu, and Xie}{Zhang
  et~al\mbox{.}}{2019}]%
        {zhang2018efficient}
\bibfield{author}{\bibinfo{person}{Rui Zhang}, \bibinfo{person}{Christian
  Walder}, \bibinfo{person}{Marian-Andrei Rizoiu}, {and}
  \bibinfo{person}{Lexing Xie}.} \bibinfo{year}{2019}\natexlab{}.
\newblock \showarticletitle{Efficient non-parametric Bayesian Hawkes
  processes}.
\newblock \bibinfo{journal}{\emph{IJCAI}} (\bibinfo{year}{2019}).
\newblock


\bibitem[\protect\citeauthoryear{Zhao, Erdogdu, He, Rajaraman, and
  Leskovec}{Zhao et~al\mbox{.}}{2015}]%
        {Zhao2015SEISMIC:Popularity}
\bibfield{author}{\bibinfo{person}{Qingyuan Zhao}, \bibinfo{person}{Murat~A.
  Erdogdu}, \bibinfo{person}{Hera~Y. He}, \bibinfo{person}{Anand Rajaraman},
  {and} \bibinfo{person}{Jure Leskovec}.} \bibinfo{year}{2015}\natexlab{}.
\newblock \showarticletitle{SEISMIC: A Self-Exciting Point Process Model for
  Predicting Tweet Popularity}. In \bibinfo{booktitle}{\emph{KDD}}.
\newblock


\end{thebibliography}

\clearpage
\appendix
\onecolumn

Accompanying the submission \textit{\titlename}.

\section{Dual Mixture Model for self-exciting processes}
\subsection{Joint Log-likelihood of Hawkes Processes}
The joint log-likelihood function of Hawkes processes given a group of cascades $\mathbb{H}$ is defined as
\begin{equation}
    \mathcal{L}(n^*, \Theta^g \mid \mathbb{H}) = \sum_{\His_i \in \mathbb{H}} \log L(n^*, \Theta^g \mid \His_i)
\end{equation}
Plugging~\cref{eq:general_likelihood} into this equation leads to
\begin{align}
    \mathcal{L}(n^*, \Theta^g \mid \mathbb{H}) &= \sum_{\His_i \in \mathbb{H}} \bracket{\sum_{t_j \in \His_{i}(T)} \log \lambda(t_j \mid \His_{i}(T)) -\int_0^T \lambda(\tau \mid \His_{i}(T)) d\tau} \\
    &= \sum_{\His_i \in \mathbb{H}} \bracket{\sum_{t_j \in \His_{i}(T)} \log \sum_{t_j \in \His_{i}(t)} n^* g(t - t_j) -\int_0^T \sum_{t_j \in \His_{i}(t)} n^* g(t - t_j) d\tau} \\
    &\stackrel{\text{(a)}}{=} \sum_{\His_i \in \mathbb{H}} \bracket{\sum_{t_j \in \His_{i}(T)} \log \sum_{t_j \in \His_{i}(t)} g(t - t_j) + \sum_{t_j \in \His_{i}(T)} \log n^* -n^* \sum_{t_j \in \His_{i}(T)} \int_{t_j}^T  g(T - t_j) d\tau} \\
    &\stackrel{\text{(b)}}{=} \sum_{\His_i \in \mathbb{H}} \bracket{\sum_{t_j \in \His_{i}(T)} \log \sum_{t_j \in \His_{i}(t)} g(t - t_j) + N_{i} \log n^* -n^* N_i} \\
    &= \mathcal{L}_g(\Theta^g \mid \mathbb{H}) + \mathcal{L}_n( n^* \mid \mathbb{H})
\end{align}
where in step (a) we separate $n^*$ due to the logarithm and we swap the order of integration and summation. Step (b) follows the assumption that $T \rightarrow \infty$.

\subsection{The Borel Mixture Model}
As the final cascade size distribution of Hawkes processes is only determined by the branching factor (\cref{sec:preliminary}), i.e. the Borel distribution, we are able to model sizes of a group of cascades as a Borel mixture model. Specifically, given a cluster number $k_v$, we aim to find the parameter set as $M^B_v = \{(n^*_1, p^B_1), \dots, (n^*_{k_v}, p^B_{k_v})\}$. The parameters are estimated via the EM algorithm following~\citepAP{tomasi2004estimating}. The log likelihood function is
\begin{equation}
    \mathcal{L}_{BMM} = \sum_{\His_{v,i} \in \mathbb{H}_v} \log \sum_{k=1}^{k_v} p^B_k \mathbb{B}(N_{v,i} \mid n^*_k)
\end{equation}
For simplicity, let $q^B(k, N_{v,i}) = p^B_k \mathbb{B}(N_{v,i} \mid n^*_k)$.
We first introduce the probability of $N_{v,i}$ being a member of $k$ which is also the E-step in the algorithm:
\begin{equation}
    p^B(k \mid N_{v,i}) = \frac{q^B(k, N_{v,i})}{\sum_{j=1}^{k_v} q^B(j, N_{v,i})}
\end{equation}
By employing Jensen's inequality, we get
\begin{align}
    \mathcal{L}_{BMM} &= \sum_{\His_{v,i} \in \mathbb{H}_v} \log \sum_{k=1}^{k_v} q^B(k, N_{v,i})\\
    &= \sum_{\His_{v,i} \in \mathbb{H}_v} \log \sum_{k=1}^{k_v} p^B(k \mid N_{v,i}) \frac{q^B(k, N_{v,i})}{p^B(k \mid N_{v,i})} \\
    & \geq  \sum_{\His_{v,i} \in \mathbb{H}_v}  \sum_{k=1}^{k_v} p^B(k \mid N_{v,i}) \log \frac{q^B(k, N_{v,i})}{p^B(k \mid N_{v,i})} \label{eq:before_q}
\end{align}
Optimizing~\cref{eq:before_q} is equivalent to optimizing the following $Q_{BMM}$ function
\begin{equation}
    Q_{BMM} = \sum_{\His_{v,i} \in \mathbb{H}_v}  \sum_{k=1}^{k_v} p^B(k \mid N_{v,i}) \log q^B(k, N_{v,i})
\end{equation}

At the \textit{Maximization} step, the parameters are updated by maximizing $Q_{BMM}$.
\begin{itemize}
    \item For updating $n^*_k$, we take the derivative of $Q_{BMM}$ w.r.t. $n^*_k$
    \begin{align}
        \frac{\partial Q_{BMM}}{\partial n^*_k} &= \sum_{\His_{v,i} \in \mathbb{H}_v}\frac{\partial   \sum_{k=1}^{k_v} p^B(k \mid N_{v,i}) \log q^B(k, N_{v,i})}{\partial n^*_k} \\
        &= \sum_{\His_{v,i} \in \mathbb{H}_v}p^B(k \mid N_{v,i})\frac{ \partial \log q^B(k, N_{v,i})}{\partial n^*_k} \\
        &= \sum_{\His_{v,i} \in \mathbb{H}_v}p^B(k \mid N_{v,i})\frac{\partial}{\partial n^*_k} \bracket{\log p^B_k \mathbb{B}(N_{v,i} \mid n^*_k)} \\
        &= \sum_{\His_{v,i} \in \mathbb{H}_v}p^B(k \mid N_{v,i}) \frac{\partial}{\partial n^*_k} \log \mathbb{B}(N_{v,i} \mid n^*_k)\\
        &= \sum_{\His_{v,i} \in \mathbb{H}_v}p^B(k \mid N_{v,i}) \frac{\frac{\partial}{\partial n^*_k} \mathbb{B}(N_{v,i} \mid n^*_k)}{\mathbb{B}(N_{v,i}\mid n^*_k)} \label{eq:lam_k_derivation}
    \end{align}
    we note that $\frac{\partial \mathbb{B}(N_{v,i} \mid n^*_k)}{\partial n^*_k}$ has a special solution
    \begin{align}
        \frac{\partial \mathbb{B}(N_{v,i} \mid n^*_k)}{\partial n^*_k} &= \frac{\partial }{\partial n^*_k} \bracket{\frac{(N_{v,i} n^*_k)^{N_{v,i}-1}e^{-N_{v,i} n^*_k}}{N_{v,i}!}} \\
        &= \frac{N_{v,i}(N_{v,i}-1)(N_{v,i} n^*_k)^{N_{v,i}-2}e^{-N_{v,i} n^*_k} -N_{v,i}(N_{v,i} n^*_k)^{N_{v,i}-1}e^{-N_{v,i} n^*_k}}{N_{v,i}!} \\
        &= \frac{\frac{N_{v,i}-1}{n^*_k}(N_{v,i} n^*_k)^{N_{v,i}-1}e^{-N_{v,i} n^*_k} -N_{v,i}(N_{v,i} n^*_k)^{N_{v,i}-1}e^{-N_{v,i} n^*_k}}{N_{v,i}!}\\
        &= \frac{N_{v,i}-N_{v,i} n^*_k - 1}{n^*_k} \mathbb{B} (N_{v,i} \mid n^*_k)
    \end{align}
    Plugging this result back to~\cref{eq:lam_k_derivation}
    \begin{align}
        \frac{\partial Q_{BMM}}{\partial n^*_k} &= \sum_{\His_{v,i} \in \mathbb{H}_v}p^B(k \mid N_{v,i}) \frac{N_{v,i}-N_{v,i} n^*_k - 1}{n^*_k}
    \end{align}
    Let the derivative be $0$ will lead to the equation
    \begin{align}
        \sum_{\His_{v,i} \in \mathbb{H}_v}p^B(k \mid N_{v,i}) (N_{v,i}-N_{v,i} n^*_k - 1) = 0
    \end{align}
    where an analytical solution exists,
    \begin{equation}
        (n^*_k)^{new} = \frac{\sum_{\His_{v,i} \in \mathbb{H}_v}p^B(k \mid N_{v,i}) (N_{v,i} - 1)}{\sum_{\His_{v,i} \in \mathbb{H}_v}p^B(k \mid N_{v,i})N_{v,i}}
    \end{equation}
    \item Updating $p^B_k$ shares same derivation steps from~\citepAP{tomasi2004estimating}
    \begin{equation}
        (p^B_k)^{news} = \frac{\sum_{\His_{v,i} \in \mathbb{H}_v}p^B(k \mid N_{v,i})}{|\mathbb{H}_v|}
    \end{equation}
\end{itemize}

Because final sizes of Hawkes processes are highly skewed towards small sizes, the estimation complexity can be reduced by counting the number of presences of various cascade sizes in $\mathbb{H}_v$, i.e., obtaining a set $C' = \{(c_i, N_{v,i})\}$ where there are $c_i$ cascades with size $N_{v,i}$. The summation over $\mathbb{H}_v$ can be then replaced by this set for efficiency.

\subsection{The Kernel Mixture Model}
We also define a mixture model for the kernel function $g(\cdot)$ (KMM) based on its likelihood function of inter-arrival times of Hawkes processes. Similarly, for a cluster number $k^v$, we denote the parameters as $M^g_v = \{(\Theta^g_1, p^g_1), \dots, (\Theta^g_{k_v}, p^g_{k_v})\}$. The log-likelihood function is
\begin{equation}
    \mathcal{L}_{KMM} = \sum_{\His_{v,i} \in \mathbb{H}_v} \log \sum_{k=1}^{k_v} p^g_k f^g( \His_i \mid \Theta^g_k)
\end{equation}
where $f^g(\His_{v,i} \mid \Theta^g) = \prod_{t_j \in \His_{v,i}}\sum_{t_z < t_j} g(t_j-t_z\mid \Theta^g)$. The membership probability (E-step) is then
\begin{equation}
    p^g(k \mid \His_{v,i}) = \frac{p^g_k f^g(\His_{v,i} \mid \Theta^g_k)}{\sum_{j=1}^{k_v} p^g_j f^g(\His_{v,j} \mid \Theta^g_j)}
\end{equation}
The function for learning parameters in EM algorithm is
\begin{equation}
    Q_{KMM} = \sum_{\His_{v,i} \in \mathbb{H}_v} \sum_{k=1}^{k_v} p^g(k \mid \His_{v,i}) \log (p^g_k f^g( \His_{v,i} \mid \theta^g_k)) = \sum_{\His_{v,i} \in \mathbb{H}_v} \sum_{k=1}^{k_v} p^g(k \mid \His_{v,i}) \log p^g_k + \sum_{\His_{v,i} \in \mathbb{H}_v} \sum_{k=1}^{k_v} p^g(k \mid \His_{v,i}) \log f^g( \His_{v,i} \mid \Theta^g_k)
\end{equation}
Updating $p^g_k$ is the same as the procedure for BMM, i.e. $(p^g_k)^{news} = \frac{\sum_{\His_{v,i} \in \mathbb{H}_v}p^g(k \mid N_{v,i})}{|\mathbb{H}_v|}$. Whereas, $\Theta^g_k$ is updated via
\begin{equation}
    (\Theta^g_k)^{new} = \argmax_{\Theta^g}  \sum_{\His_{v,i} \in \mathbb{H}_v} p^g(k\mid \His_{v,i}) \log f^g( \His_{v,i} \mid \Theta^g)
\end{equation}
As there is no analytical solution for the power-law kernel, we solve $(\Theta^g_k)^{new}$ with a non-linear solver, Ipopt~\citep{Wachter2006}.

\section{Additional results on \textit{ActiveRT2017} and \textit{RNCNIX}}
\subsection{Inter-arrival times of cascades}
\begin{figure}[H]
	\centering
	\begin{subfigure}{.35\textwidth}
		\includegraphics[width=\textwidth,page=1]{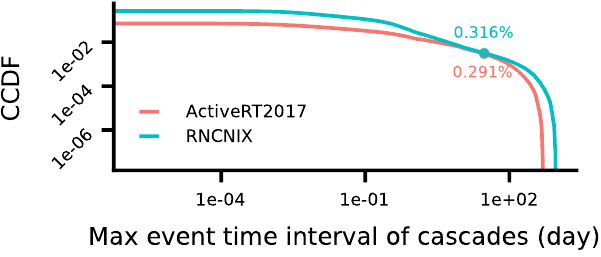}
		\vspace{-6mm}
	\end{subfigure}
	\caption{
		The complementary cumulative density function (CCDF) of the maximum inter-arrival time of cascades from \textit{ActiveRT2017} and \textit{RNCNIX}. The labeled points show the proportions of cascades with inter-arrival times larger than \textbf{30} days. The CCDF curves does not start at $1$ as maximum inter-arrival times for single-event cascades are considered $0$ thus being filtered during the log transformation of x axis.
	}
	\label{fig:dataset_interarrival}
	\vspace{-2mm}
\end{figure}

\cref{fig:dataset_interarrival} shows the complementary cumulative density function of the inter-arrival times of cascades from the two datasets. This shows that our assumption --- cascades that do not get new retweet for $30$ days are finished --- accounts for more than $99\%$ of cascades in our datasets.

\subsection{Weighted density plots of $n^*$ and $\theta$}
In addition to \cref{fig:param_density}, we show here weighted density plots of $n^*$ and $\theta$ of the fitted parameters on \acth{} and \fakeh{}, where the density weights refer to the mixture components weights.
\begin{figure}[H]
    \centering
    \begin{subfigure}{0.495\textwidth}
        \includegraphics[width=\textwidth,page=3]{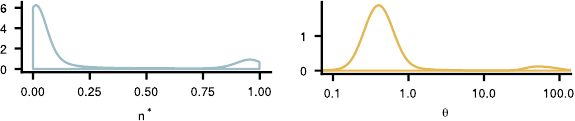}
        \caption{\acth{}}
    \end{subfigure}
    \begin{subfigure}{0.495\textwidth}
        \includegraphics[width=\textwidth,page=3]{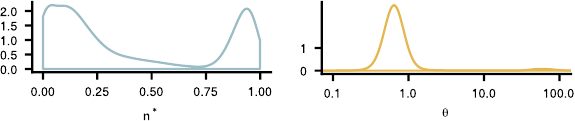}
        \caption{\fakeh{}}
    \end{subfigure}
	\caption{
        Weighted density plots of content \revA{virality} $n^*$  of BMMs and content influence decay $\theta$ of KMMs fitted on two datasets.
    }
    \label{fig:parameter_measurement2}
\end{figure}

\subsection{Analysis of mixture models on explaining popular cascades and unpopular cascades}
In this section, we first choose an online items with its fitted dual mixture model and then show the posterior mixture component assignments of individual cascades relating to this item. A YouTube video (ID: \textit{QvCj3wsXQDQ}) is chosen and the following figure shows cascades are assigned to $5$ different components. Overall, this figure shows that unpopular and popular cascades are modeled by different mixture components which reinforces the assumption that the proposed dual mixture model leverages the information from unpopular diffusion cascades.

\begin{figure}[H]
    \centering
    \begin{subfigure}{0.5\textwidth}
        \includegraphics[width=\textwidth]{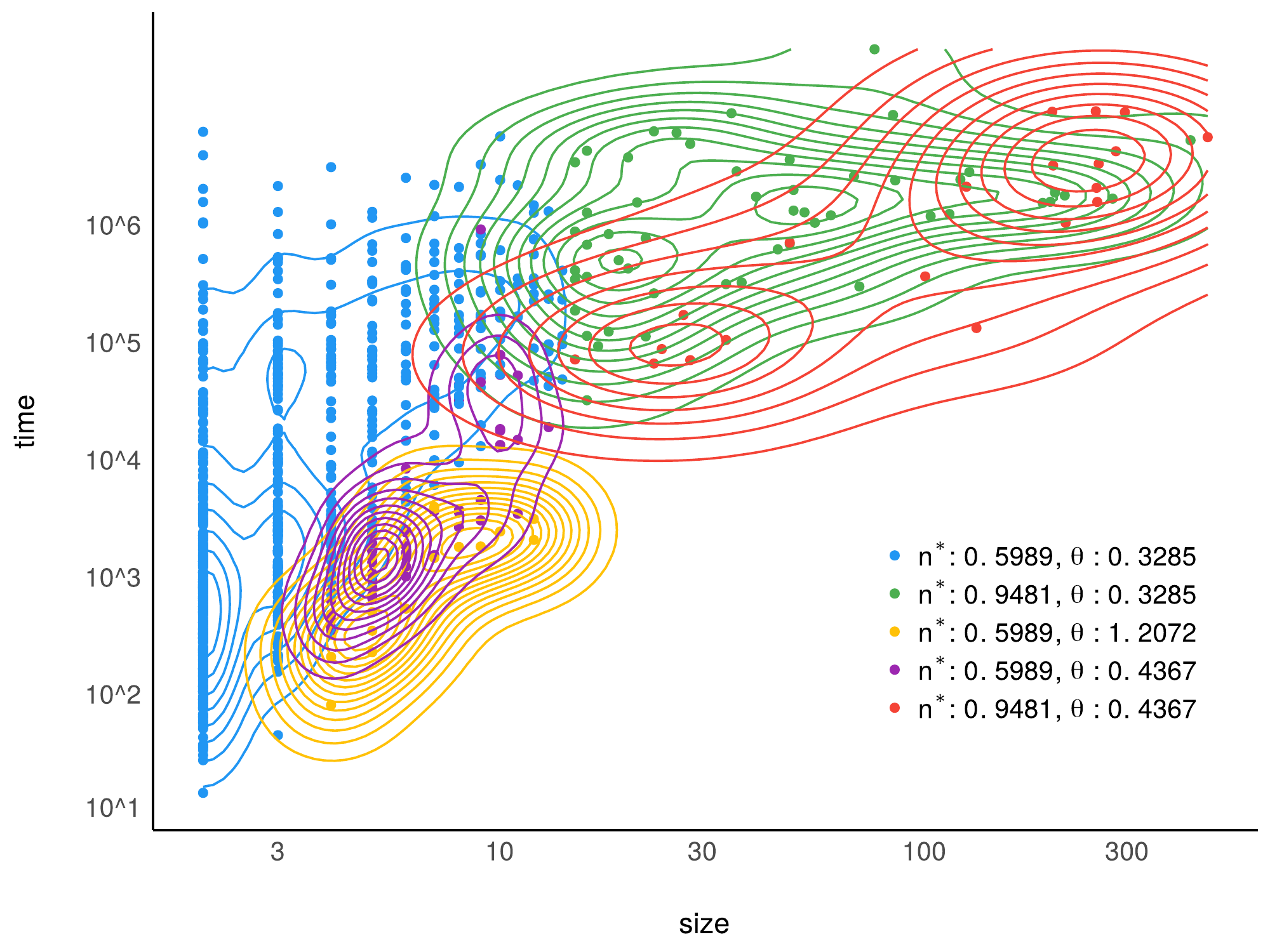}
    \end{subfigure}
	\caption{
        Posterior assignments to the fitted mixture components of cascades relating to a YouTube video (ID: \textit{QvCj3wsXQDQ}). Each dot is a diffusion cascade positioned by the final cascade size (x axis) and the total diffusion time (y axis). Each color represents a mixture component whose parameters are shown in the legends.
    }
\end{figure}

\subsection{Category-level measurement}
\cref{fig:parameter_measurement2} quantifies online items from \acth{} at the category level in the same form as~\cref{fig:measurement_b}.
\begin{figure}[H]
    \centering
    \begin{subfigure}{0.7\textwidth}
        \includegraphics[width=\textwidth]{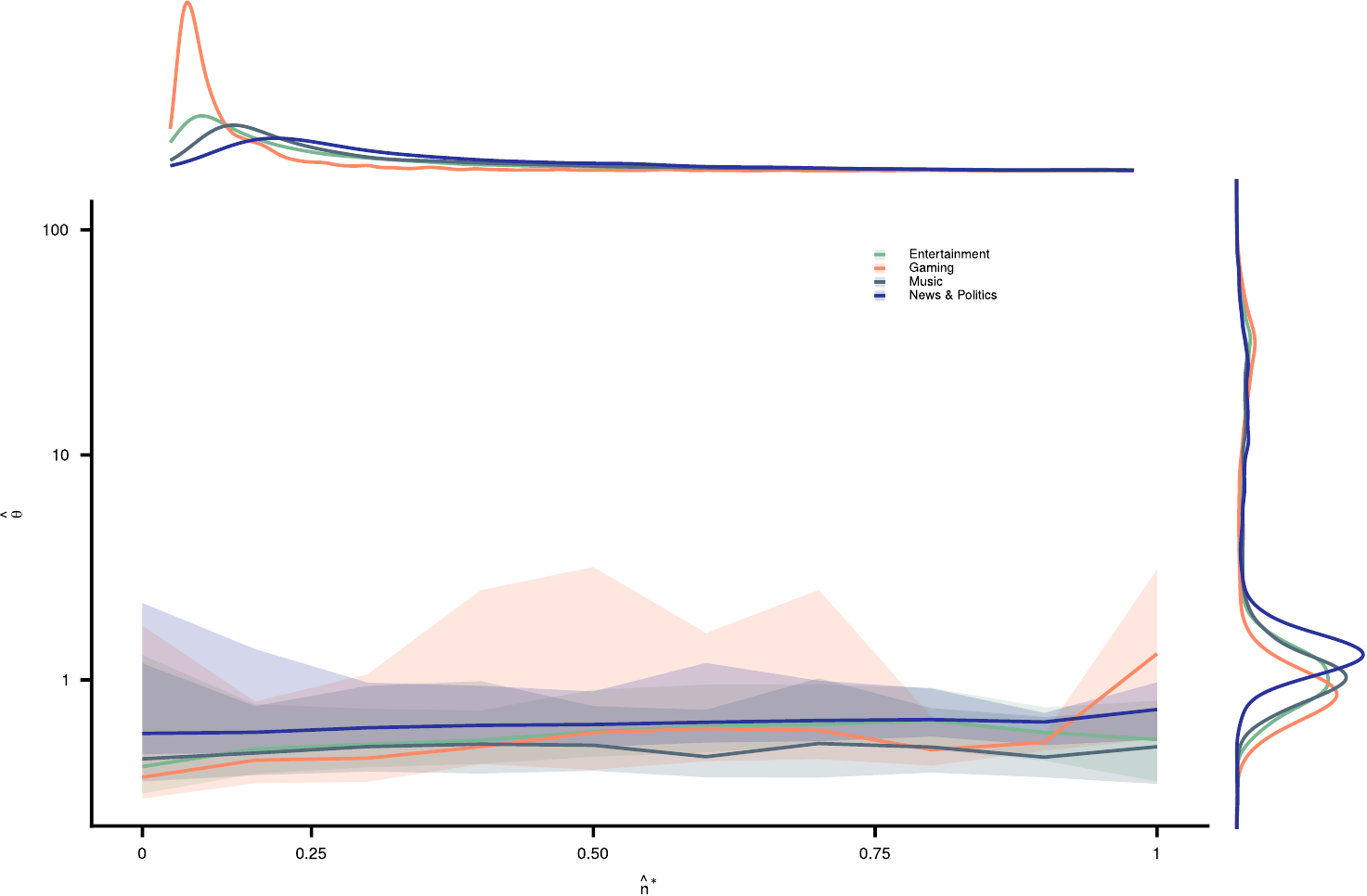}
    \end{subfigure}
	\caption{
        Quantify online items at the category level via the aggregated model parameters, $\hat{n}^*$ and $\hat{\theta}$ of \acth{}. Four popular YouTube video categories, \textit{Music}, \textit{Entertainment}, \textit{Gaming} and \textit{News \& Politics} from \acth{}: the median and $25\%$/$75\%$ quantiles of $\hat{\theta}$ (y axis) at varying $\hat{n}^*$ values (x axis) are presented along with densities of $\hat{n}^*$ and $\hat{\theta}$ by sides.
    }
    \label{fig:parameter_measurement2}
\end{figure}

\subsection{Comparing mixture models to non-mixture models on distinguishing publishers}
\begin{figure}[H]
    \centering
    \begin{subfigure}{0.3\textwidth}
        \includegraphics[width=\textwidth]{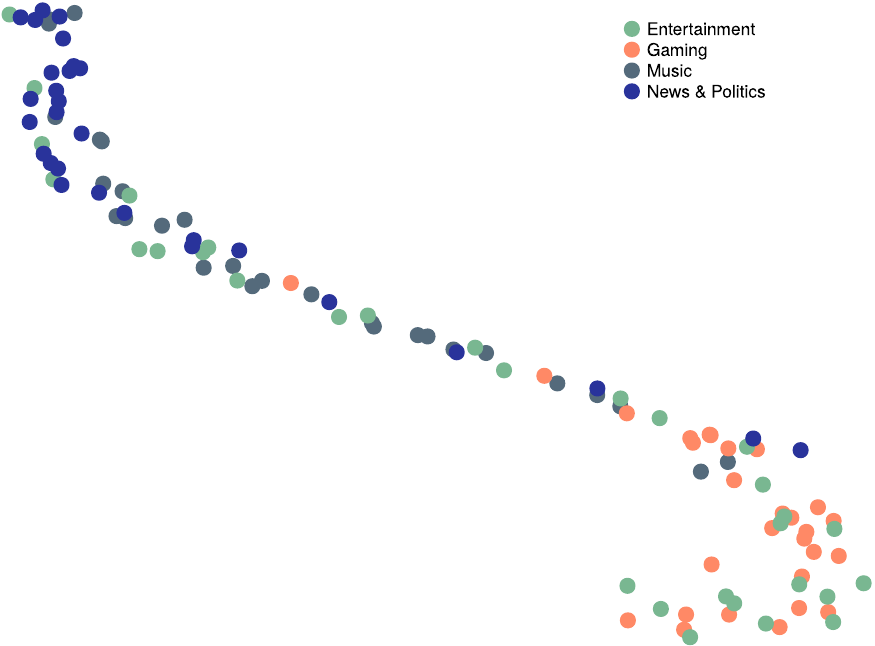}
        \caption{\acth{}}
    \end{subfigure}
    \hspace{5mm}
    \begin{subfigure}{0.3\textwidth}
        \includegraphics[width=\textwidth]{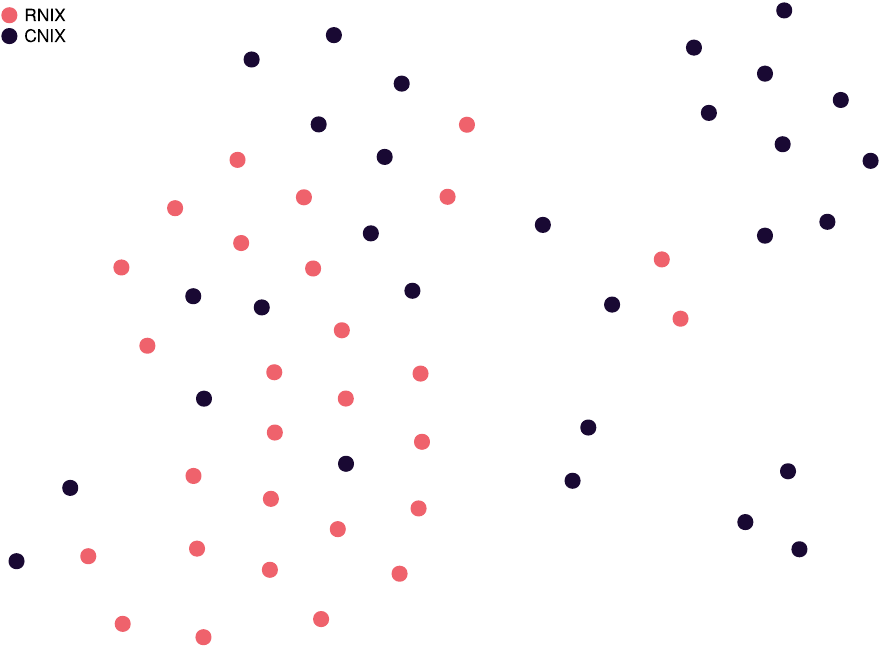}
        \caption{\fakeh{}}
    \end{subfigure}
	\caption{
        Clustering of publishers with parameters fitted on all cascades related to all online items from a given publisher.
    }
    \label{fig:clustering_2}
\end{figure}

The non-mixture models are individual power-law decayed Hawkes processes fitted jointly on all cascades related to all online items from a given publisher. Using the fitted parameters $[n^*, c, \theta]^\mathsf{T}$ of each publisher, we use t-SNE~\citepAP{maaten2008visualizing} to clustering the publishers as in~\cref{fig:clustering_2}. In comparison, \cref{fig:channel_clustering} depicts a better separability as the diffusion embeddings are applied which encode more diverse item-level temporal information.

\subsection{Forecasting for unseen content}
\begin{figure}[H]
    \centering
    \begin{subfigure}{0.95\textwidth}
        \includegraphics[width=\textwidth,page=1]{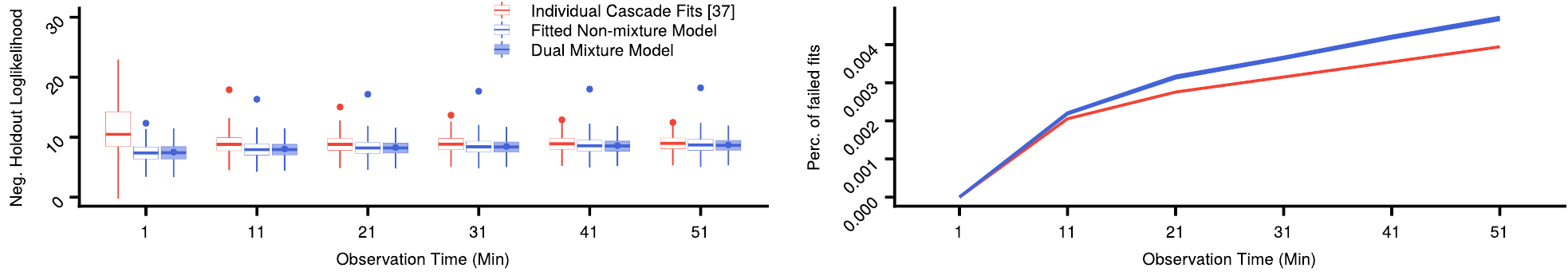}
        \caption{On cascades with less than 50 events}
        \label{subfig:hll2-pop}
    \end{subfigure}

    \begin{subfigure}{0.95\textwidth}
        \includegraphics[width=\textwidth,page=2]{images/cikm_experiment_hll2}
        \caption{On cascades with more than 50 events}
        \label{subfig:hll2-unpop}
    \end{subfigure}
	\caption{
		\textbf{Negative holdout log-likelihood per event} (left panels) and \textbf{percentages of failed cascades} (right panels) computed from the fitted non-mixture model, the dual mixture models and~\citep{Mishra2016FeaturePrediction} on different subsets of \actc{} given different observation times --- lower is better.
	}
	\label{fig:hll2}
	\vspace{-2mm}
\end{figure}

We present here the results on evaluating the generalization performance and the item popularity predictions on \actc{} and \fakec{}. In \cref{fig:hll2}, we further split \actc{} into two parts based on cascade popularities to compare varying performances of different models. We observe from \cref{subfig:hll2-pop} that on popular cascades, both mixture and non-mixture models outperform the benchmark~\citep{Mishra2016FeaturePrediction} with similar percentages of failed cascades for all models. This indicates that joint fittings from historical cascades provide the most performance gain on popular cascades. However, in \cref{subfig:hll2-unpop}, we note that the proposed dual mixture model achieves the best negative holdout likelihood values among the three. Most notable, as much less events are available for learning, much higher proportions of failed fits are shown for individual cascade fits~\citep{Mishra2016FeaturePrediction}.

\begin{figure}[H]
    \centering
    \begin{subfigure}{0.4\textwidth}
        \includegraphics[width=\textwidth]{images/cikm_experiment_hll}
        \vspace{-5mm}
        \caption{\actc{}}
        \vspace{-4mm}
        \label{subfig:hll-PL}
    \end{subfigure}
    \hspace*{5mm}
    \begin{subfigure}{0.4\textwidth}
        \includegraphics[width=\textwidth]{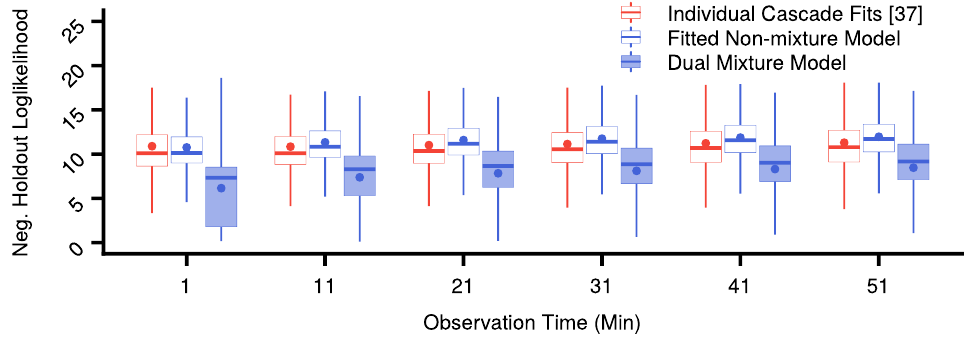}
        \vspace{-5mm}
        \caption{\fakec{}}
        \vspace{-4mm}
        \label{subfig:hll-EXP}
    \end{subfigure}
	\caption{
		\textbf{Negative holdout log-likelihood per event} computed from the fitted non-mixture model, the dual mixture models and~\citep{Mishra2016FeaturePrediction} on \actc{} and \fakec{} given different observation times --- lower is better.
	}
	\label{fig:hll}
\end{figure}

\begin{figure}[H]
	\centering
	\begin{subfigure}{0.485\textwidth}
		\includegraphics[width=\textwidth,page=1]{images/cikm_experiment_size_prediction_PL_fullMM.pdf}
		\vspace{-5mm}
        \caption{\actc}
		\vspace{-3mm}
	\end{subfigure}
	\hspace*{\fill}
	\begin{subfigure}{0.485\textwidth}
		\includegraphics[width=\textwidth,page=1]{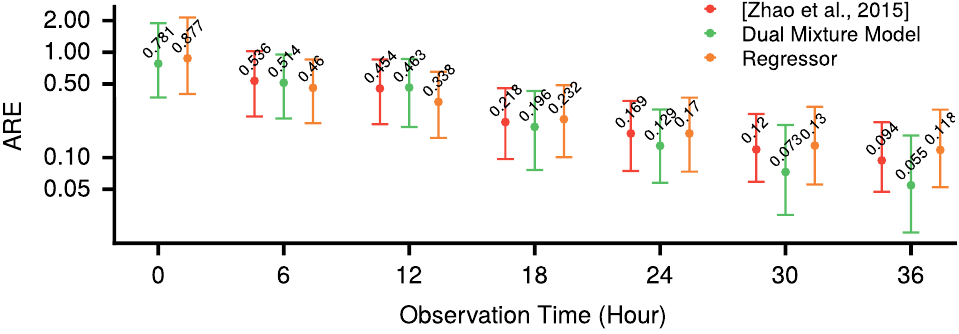}
		\vspace{-5mm}
        \caption{\fakec}
		\vspace{-3mm}
	\end{subfigure}
	\caption{
        Newly published online item final popularity predictions of three models on \actc and \fakec, evaluated with Absolute Relative Error --- lower is better. Times at x axis are the observation time since an online item was published. The dots indicate the median values and error bars give the first/third quarters of the ARE values.
	}
	\label{fig:prediction}
	\vspace{-5mm}
\end{figure}

\bibliographystyleAP{ACM-Reference-Format}
\bibliographyAP{acm}
\end{document}